\documentclass[sn-mathphys-num]{sn-jnl}% Math and Physical Sciences Numbered Reference Style 
\usepackage{mathtools}
\usepackage{comment}
\usepackage{subcaption}
\usepackage{stackengine} 
\usepackage{adjustbox}
\usepackage{float}
\usepackage{multirow}%
\usepackage{amsmath,amssymb,amsfonts}%
\usepackage{amsthm}%
\usepackage{mathrsfs}%
\usepackage[title]{appendix}%
\usepackage{xcolor}%
\usepackage{textcomp}%
\usepackage{manyfoot}%
\usepackage{booktabs}%
\usepackage{algorithm}%
\usepackage{algorithmicx}%
\usepackage{algpseudocode}%
\usepackage{listings}%
\usepackage{subcaption}
\usepackage{graphicx} 
\usepackage{adjustbox} 
\usepackage{dblfloatfix}
\usepackage{float}
\usepackage{braket}
\usepackage[normalem]{ulem}
\usepackage{bbm}
\usepackage{pst-all}
\usepackage{pst-node}
\usepackage{pstricks-add}
\usepackage{pst-optexp}
\usepackage{pst-coil}
\usepackage{auto-pst-pdf}
\usepackage{tikz}
\usetikzlibrary{calc,shapes.geometric,arrows,positioning,intersections}
\usepackage{graphicx}
\usepackage{hyperref}
\usepackage{xcolor}
\usepackage{matlab-prettifier}
\usepackage{caption}
\usepackage{listings}
\usepackage{xcolor}

\lstdefinestyle{matlabstyle}{
    language=Matlab,
    basicstyle=\ttfamily\small,
    keywordstyle=\color[RGB]{108, 142, 191}, % Soft blue for keywords
    commentstyle=\color[RGB]{153, 153, 153}, % Light gray for comments
    stringstyle=\color[RGB]{209, 139, 71}, % Soft orange for strings
    numbers=left,
    numberstyle=\tiny\color[RGB]{102, 102, 102}, % Medium gray for line numbers
    breaklines=true,
    frame=single,
    captionpos=b,
    showspaces=false,
    showstringspaces=false,
    tabsize=4,
    backgroundcolor=\color[RGB]{245, 245, 245} % Very light gray background
}
\theoremstyle{thmstyleone}%
\theoremstyle{thmstyletwo}%

\theoremstyle{thmstylethree}%

\begin{document}

\title[Article Title]{Dense Associative Memory in a Nonlinear Optical Hopfield Neural Network}


\author[1,2]{\fnm{Khalid} \sur{Musa}}\email{kmusa@stevens.edu}

\author[1,2]{\fnm{Santosh} \sur{Kumar}}\email{skumar5@stevens.edu}

\author[1,2]{\fnm{Michael} \sur{Katidis}}%\email{mkatidi1@stevens.edu}

\author*[1,2]{\fnm{Yu-Ping} \sur{Huang}}\email{yhuang5@stevens.edu}


\affil[1]{Department of Physics, Stevens Institute of Technology, Hoboken, New Jersey 07030, USA}
\affil[2]{Centre for Quantum Science and Engineering, Stevens Institute of Technology, Hoboken, New Jersey 07030, USA}


\abstract{Modern Hopfield Neural Networks (HNNs), also known as Dense Associative Memories (DAMs), enhance the performance of simple recurrent neural networks by leveraging the nonlinearities in their energy functions. They have broad applications in combinatorial optimization, high-capacity memory storage, deep learning transformers,  and correlated pattern recognition. Thus far, research on DAMs has been primarily theoretical, with implementations limited to CPUs and GPUs. In this work, for the first time to our knowledge, we propose and experimentally demonstrate a nonlinear optical Hopfield neural network (NOHNN) system for realizing DAMs using correlated patterns. Our NOHNN incorporates effective 2-body and 4-body interactions in its energy function. The inclusion of 4-body interaction scores a minimum ten-fold improvement in the number of uncorrelated patterns that can be stored and retrieved, significantly surpassing the traditional capacity limit for traditional HNNs. For correlated patterns, depending on their average correlation, up to 50 times more patterns can be stored compared to traditional HNNs. To test the system's robustness, the benchmark testing is performed on MNIST handwritten digit patterns. The results show a 5.5 times improvement in the pattern storage along with the retrieval of cleaner and less noisy patterns. These results highlight the potential of nonlinear optical DAMs for practical applications in challenging big-data optimization, computer vision and graph network tasks. 
}




\maketitle 

\section{Introduction}\label{sec1}

Hopfield Neural Networks (HNNs) are a class of recurrent neural networks (RNNs) inspired by the Ising models in statistical physics to simulate the structure of biological neurons and principles of human cognitive psychology \cite{katz1996intrinsic, katz1994dynamic,Schurr2024}. These networks serve as versatile models for associative memory \cite{hopfield1982neural} and pattern recognition \cite{suganthan1995pattern, nasrabadi1991object}, bridging the gap between neuroscience and machine learning. One of the key strengths of HNNs lies in their ability to store and retrieve patterns across the entire network, mimicking the operation of human memory. This property makes them particularly useful in applications such as content-addressable memory systems and optimization problems \cite{tank1986simple,hopfield1985neural}. By leveraging associative memory dynamics, HNNs reconstruct missing information, akin to how masked autoencoders use machine learning to infer masked portions of data \cite{he2022masked}. Furthermore, HNNs exhibit remarkable tolerance to noise, enabling reliable pattern retrieval even in the presence of incomplete or corrupted inputs \cite{agliari2020tolerance,alonso2024sparse,hu2015associative}, making them valuable for applications in error-correcting codes \cite{sourlas1989spin}. Their energy-based framework also ensures stability and reliability in diverse tasks, including optimization and image restoration \cite{paik1992image}. \bigskip

Recent efforts to improve energy efficiency in large-scale deep learning \cite{6757323} have led to innovations like optical processors, which reduce energy consumption while maintaining high performance (see  \cite{Wetzstein2020,shastri2021photonics} and references therein). These advancements, initially applied to Transformer-based architectures \cite{wang2024optical, anderson2023optical}, could also enhance the scalability and efficiency of HNNs. Moreover, by integrating modern techniques such as deep learning \cite{leonetti2024photonic,kumar2021robust} and hybrid architectures \cite{xu2024large, ramsauer2020hopfield, krotov2020large,denz2013optical}, the utility of HNNs can be expanded to more complex domains, including natural language processing, recommendation systems, and even quantum computing \cite{bausch2020quantum, marsh2021enhancing}. Despite their early success, HNNs and Ising models face challenges such as limited storage capacity \cite{negri2023storage} and susceptibility to spurious states \cite{amit1985storing}. These challenges can be ascribed to their reliance on global energy minimization, where the system converges to the absolute lowest-energy state, often limiting their flexibility in handling complex or correlated patterns. \bigskip

A promising approach to expand the HNNs' utilities is by incorporating higher-order nonlinear interactions, which gives rise to dense associative memories (DAMs) \cite{krotov2020large, demircigil2017model}. These modern HNNs leverage multi-body interactions in their energy functions to exponentially increase storage capacity and robustness by converging to the nearest local minima rather than the the global minimum \cite{ramsauer2020hopfield, bausch2020quantum, lucibello2024exponential}. This shift has expanded their applicability to complex tasks such as natural language processing, computer vision, and combinatorial optimization, bridging the gap between traditional associative memory and modern deep learning architectures \cite{li2024first, cai2020power}. So far, development on DAMs has been primarily theoretical, with implementations limited to CPUs and GPUs. While these implementations have demonstrated the potential of DAMs for pattern storage and retrieval, their scalability remains constrained by the computational and memory limitations of traditional digital hardware \cite{krotov2016dense,krotov2020large}. Additionally, emerging approaches, such as optical and quantum-inspired implementations, offer promising alternatives for overcoming some of these bottlenecks by leveraging high-dimensional encoding, nonlinearity, and massive parallelism \cite{marsh2021enhancing, fan2023photonic,mcmahon2023physics,Wu2025}. Exploring DAMs beyond traditional digital architectures could pave the way for more efficient, large-scale associative memory systems.\bigskip

In our previous work, we demonstrated robust pattern retrieval in a traditional optical Hopfield neural network (OHNN) using the Ising type model (without the use of optical nonlinearity) \cite{katidis2024robust}. We observed a phase transition in storage capacity, confirming $0.138$ as the critical threshold ratio of the pattern number to the neuron number, and showed efficient retrieval even with average random masking up to 25\%, highlighting the system’s high tolerance to perturbations below the critical capacity. In this paper, we present the first experimental demonstration of DAMs using a nonlinear optical Hopfield neural network (NOHNN) employing correlated patterns. By incorporating both 2-body and 4-body interactions in the energy function, our NOHNN achieves more than a tenfold improvement in the critical number of storable and retrievable uncorrelated patterns, surpassing the traditional capacity limit. For correlated patterns, the improvement scales with the average correlation, surpassing the traditional limit by up to 50 times. A benchmark test on MNIST handwritten digit patterns further confirms the scalability and robustness of our approach. These findings highlight the potential of nonlinear optical DAMs for real-world applications in complex optimization, computer vision, and graph network analyses. Our results not only help advance photonic computing but also demonstrate its viability as an energy-efficient alternative for high-capacity memory storage and a pathway to simplifying deep-learning architectures for advanced AI applications \cite{spens2024generative,wang2022optical}. In the future, the NOHNN can be extended to sub-attojoule optics, reducing the per-calculation energy cost to below yocto joules and leveraging quantum superposition and interference to unlock unprecedented advantages in speed, capacity, and energy efficiency  \cite{leonetti2024photonic,li2024first,shen2017deep,cai2020power,ma2025quantum}. 

\section{Theoretical Background}

This section provides the theoretical foundation for the models and relevant techniques. We begin by introducing the traditional Hopfield neural network and its extension to DAM, which incorporates higher-order interactions to improve storage capacity and robustness. We then discuss the NOHNN, a framework that leverages nonlinear optical techniques to overcome the computational bottlenecks of traditional and DAM-based Hopfield networks, enabling efficient large-scale computations. 

\subsection{Dense Associative Model}
The traditional HNN consists of a fully connected array of $N$ neurons \( \boldsymbol{ \sigma} =\) \{$\sigma_1, \sigma_2, \dots, \sigma_N$\}, where each neuron $\sigma_i$ is connected to every other neuron $\sigma_j$ with a connection weight $w_{ij}$. Each neuron has a binary state; it is considered \textquotedblleft on" when $\sigma_i = +1$, and \textquotedblleft off" when $\sigma_i = -1$. The collective dynamics of the network can be described using an energy-based model, with the system's total energy given by: 

\begin{equation}
    H = - \sum_{i,j}^{N} w_{ij} \sigma_i \sigma_j.
    \label{equation1}
\end{equation}

\noindent  The weight term can be defined as $w_{ij} = \sum_{\mu = 1}^{K} \xi_{i}^{\scriptscriptstyle(\mu)} \xi_{j}^{\scriptscriptstyle(\mu)}$, to introduce $K$ stored binary memories or patterns into our system, giving:

\begin{equation}
    H = - \sum_{\mu = 1}^{K} (\boldsymbol{\xi^{\scriptscriptstyle(\mu)}} \cdot \boldsymbol{ \sigma})^2.
    \label{equation2}
\end{equation}

This formulation defines the network as an associative memory (AM) model, where \(\boldsymbol{\xi^{\scriptscriptstyle(\mu)}}\) represents the $\mu$-\text{th} stored pattern, with $\mu=1,2,\dots,K$ \cite{katidis2024robust}. In ideal cases, local minima in the energy landscape correspond to stored patterns \(\ \boldsymbol{\sigma} = \boldsymbol{\xi^{\scriptscriptstyle(\mu)}}\), though spurious minima may arise in practice. While traditional HNNs laid the groundwork, their susceptibility to spurious states and cross-talk motivated the development of models incorporating multi-neuron interactions \cite{koh2009increasing}, which reshape the energy landscape to improve pattern stability and discrimination. This approach is described by the DAM with the following energy function:

%In DAM, this is achieved by incorporating higher-order interactions between neuron states. 
\begin{equation}
    \tilde{H} = - \sum_{\mu = 1}^{K} F(\boldsymbol{\xi^{\scriptscriptstyle(\mu)}} \cdot \boldsymbol{ \sigma}),
    \label{equation3}
\end{equation}

\noindent where \(F(x)\) is a general nonlinear function. A common function used is \(F(x) = x^n\), with \(n\) denoting the order of interaction \cite{krotov2016dense,bao2022capacity}. The \(n\)-body interactions specify the number of neuron states interacting simultaneously, allowing the DAM model to store a significantly larger number of patterns. The steepness of the local minima in the energy landscape is also increased, leading to faster convergence and reduced cross-talk between stored patterns compared to traditional HNNs with only 2-body interaction \cite{krotov2016dense}. This leads to greater storage capacity and robustness in the model. To test the robustness of DAM in pattern recall, a pattern \(\boldsymbol{\xi^{\scriptscriptstyle(\mu)}}\) is blurred or masked by altering a $\delta$-fraction of its values. In this work, it is done by randomly flipping \(\delta \times N\) neurons resulting in a masked pattern, \(\boldsymbol{\xi_{\text{mask}}}\), which is used as the initial neuron structure \((\boldsymbol{ \sigma}(t=0) = \boldsymbol{\xi_{\text{mask}}})\). 

%This configuration is designed to be within range of the local minima of the respective stored pattern \(\boldsymbol{\xi^{\scriptscriptstyle(\mu)}}\). (ADD somewhere else)

\subsection{Retrieval Accuracy \& DAM Limitations}
\label{Retrieval accuracy}

The DAM run begins by initializing the neuron configuration and comparing it with each original stored pattern. The higher the similarity between the two, the lower the energy value gets. A neuron's value is then randomly altered and the similarity testing is then repeated to find the new energy of the system. This process adopts the Monte-Carlo algorithm which updates the neurons iteratively by decreasing the system's total energy \(H\). After $T$ iterations, the network evolves to a final configuration \(\boldsymbol{ \sigma}(t=T) = \boldsymbol{r}\), referred to as the replica. Ideally, the replica reproduces the original unmasked pattern \(\boldsymbol{\xi^{\scriptscriptstyle(\mu)}}\). This procedure tests the network's ability to recall and reconstruct perturbed stored patterns. Note that if the masking percentage \(\delta > 0.5\), the neuron configuration is more likely to converge to the inverted pattern, \(-\boldsymbol{\xi^{\scriptscriptstyle(\mu)}}\). This occurs due to the symmetry in the energy function between the original and inverted patterns.  \bigskip

To assess the network’s ability to retrieve any stored pattern, we conduct \( K\) runs, each testing a different masked pattern. Together, these runs form a single trial. So, in the \(\nu\)-th run of a trial, the \(\nu\)-th stored pattern \(\boldsymbol{\xi^{\scriptscriptstyle(\nu)}}\) is masked and tested to retrieve replica $\boldsymbol{r^{\scriptscriptstyle(\nu)}}$ with $\nu = 1,2,\dots,K$. To quantify the success of the recall process for a single trial, we compute the overlap matrix \(\boldsymbol m\) to compare all replicas against all stored patterns:
\begin{equation}
    m_{\mu \nu} = \frac{1}{N} \, \boldsymbol{\xi^{\scriptscriptstyle(\mu)}} \cdot \boldsymbol{r^{\scriptscriptstyle (\nu)}}.
    \label{equation4}
\end{equation}

In the ideal case of perfect recall, \(\boldsymbol{m}\) will be diagonal, with \(m_{\nu\nu} = 1\) and \(m_{\mu \nu} \approx 0\) for \(\mu \neq \nu\). This diagonal structure arises because the replica of the \(\nu\)-th run will have maximum similarity with the \(\nu\)-th stored pattern and negligible similarity with all other patterns. In practice, however, correlations between patterns or imperfect recall may result in significant non-zero off-diagonal elements. A recall is considered successful if the diagonal element \(m_{\nu \nu}\) of the overlap matrix is (a) the maximum in its column and (b) exceeds all off-diagonal elements \(\boldsymbol{m_{\mu \nu}
}\) (\(\mu \neq \nu\)) by a distance of \(\tau \times m_{\nu \nu}\) to account for spurious minima, where \(\tau\) is a similarity threshold constant that depends on the type of pattern (e.g., \(\tau = 0.60\) for orthogonal patterns and \(\tau = 0.20\) for correlated patterns). \bigskip

The retrieval accuracy \(\eta\), which measures the network's ability to correctly recall a set of stored patterns, is calculated as the average proportion of successful recalls across $L$ trials. When $\eta <1$, the network fails to reliably retrieve all $K$ patterns, indicating that $K$ has surpassed its critical pattern capacity $K_c$. Hence, $K_c$ is the maximum number of patterns that can be stored while maintaining $\eta = 1$. Traditional AM models operate with 2-body interaction (\(n=2\)), whose theoretical storage limit is $K_c=\alpha_c N$, with \(\alpha_c = 0.138\) \cite{mceliece1987capacity,hopfield1982neural}. This limitation motivates the exploration of models with higher-order interactions, as \(K_{c}\) scales exponentially with \(n\). For example, a 4-body interaction model gives \(K_{c} =\alpha_c^{(4)} N^{3}\), where \(\alpha_c^{(4)}\) is the \(4\)-body critical storage capacity \cite{de2023effect}. \bigskip

%Another factor arises from the system's order of interaction, \(n\).

The above assumes an ideal operation. In practice, however, the achievable $K_c$ is limited by several factors. The first is the masking, where as its percentage $\delta$ increases, the correct configuration becomes more difficult to distinguish from the others, causing incorrect recall. Consequently, there is an inverse relationship between \(K_{c}\) and \(\delta\), where $K_c$ decreases quadratically on $\delta$ \cite{bao2022capacity}. The second is the correlation among patterns. Highly correlated patterns share similarities, which can lead to increased interference during retrieval. As such, the network can confuse one pattern for another. This effect becomes more pronounced as the number of patterns \(K\) increases. Together, these two factors define the practical limits of the network's memory capacity and guide the design of more robust associative memory models.

%will be explored in detail, 

%More in-depth discussion on the correlation used in this paper is provided in Appendix \ref{Correlation in Patterns}. $\boldsymbol{\rho}$, describing the similarities between the stored patterns, plays a significant role in determining \(K_c\).  

%The interplay between the masking percentage \(\delta\), the order of interaction \(n\), and the pattern correlation will be explored in detail in subsequent sections. Together,

\subsection{Nonlinear Optical Hopfield Neural Network (NOHNN)}
Building on the DAM framework---which enhances storage capacity through high-order interactions at the cost of increased computational complexity---the nonlinear optics-based OHNN exploits the intrinsic parallelism of optical systems to alleviate this bottleneck. In digital implementations, as the number of neurons grows, the processing speed of traditional systems slows down significantly, and matrix multiplications become increasingly time-consuming. The OHNN overcomes these limitations by leveraging optical techniques, enabling high-speed, parallel computations that are infeasible with digital systems \cite{zhou2022photonic,katidis2024robust}. 

\begin{figure}[H]
    \centering
    %\hspace*{-10mm}
    \includegraphics[width=1\linewidth]{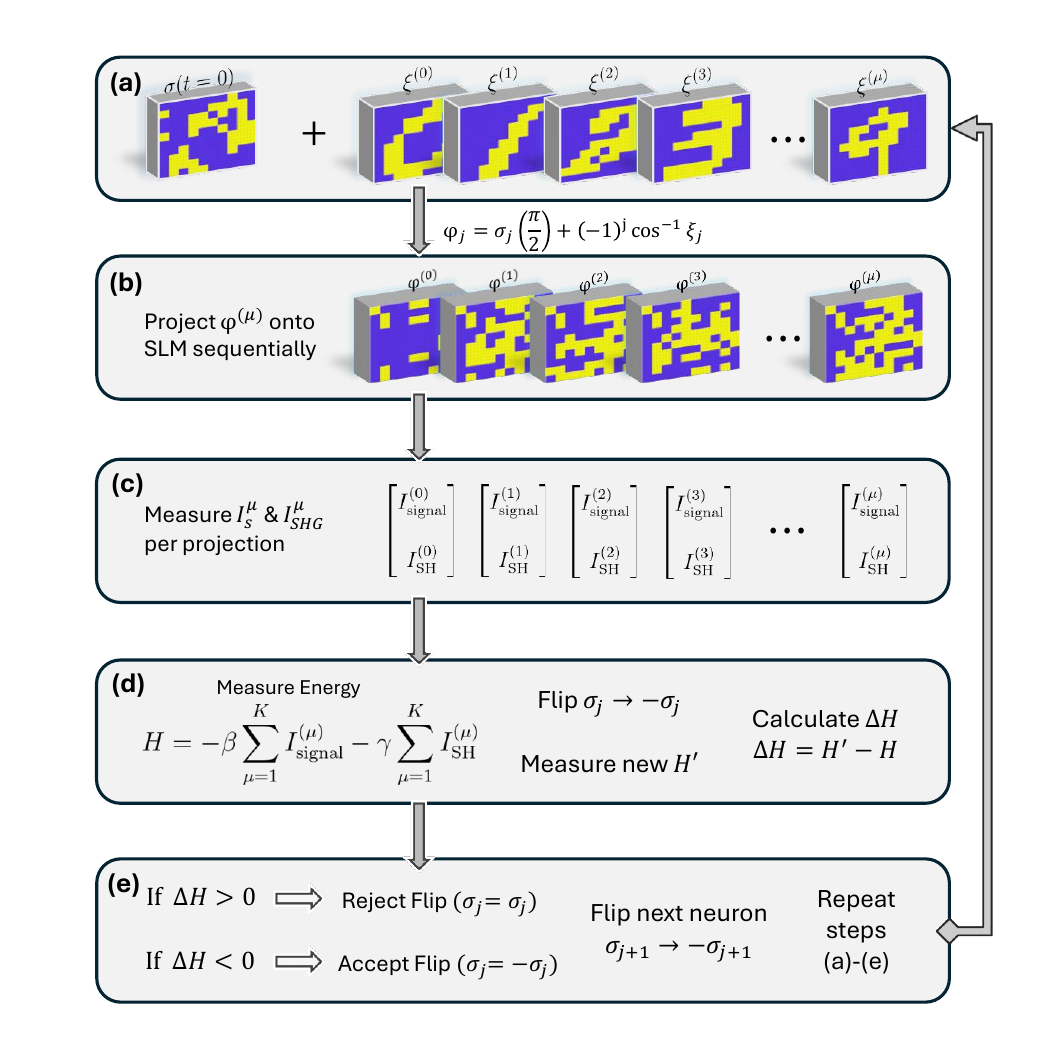}
    \caption{{\bf Workflow for Implementing the NOHNN.} (a) Gauge transformation \cite{fang2021experimental,katidis2024robust} is used to create new phase patterns $\varphi^{\scriptscriptstyle(\mu)}$. (b) Phase patterns are sequentially projected into the SLM. (c) Intensities of modulated beams are measured per projection. (d) System energy $H$ is computed after all $K$ projections, followed by trial spin flip and energy $H'$ measurement. (e) Energy difference $\Delta H = H' - H$ is evaluated and energy minimization determines flip acceptance. The process is repeated until $t=T$ iterations.}
    \label{fig:OHNN-Drawing}
\end{figure}

For a DAM network with \(K\) patterns, the Hamiltonian (i.e., total energy) describing the system's dynamics is given by:

\begin{equation}
H = - \beta \sum_{\mu = 1}^{K} (\boldsymbol{\xi^{\scriptscriptstyle(\mu)}} \cdot \boldsymbol{ \sigma})^2 - \gamma \sum_{\mu = 1}^{K} (\boldsymbol{\xi^{\scriptscriptstyle(\mu)}} \cdot \boldsymbol{ \sigma})^4,
    \label{equation5}
\end{equation}

\noindent
where \(\beta\) and \(\gamma\) are coefficients scaling the contributions of 2-body and 4-body interactions, respectively. To address the computational challenges associated with large \(N\), the above Hamiltonian can be optically implemented to allow for instantaneous matrix calculations through parallel linear optical processing \cite{kumar2020large,pierangeli2019large,denz2013optical}. Then, the resultant beam is passed through a nonlinear-optical medium to produce the above nonlinear terms \cite{wang2024large}. As derived in Appendix \ref{theoretical}, \((\boldsymbol{\xi^{\scriptscriptstyle(\mu)}} \cdot \boldsymbol{ \sigma})^2\) and \( (\boldsymbol{\xi^{\scriptscriptstyle(\mu)}} \cdot \boldsymbol{ \sigma})^4\) in Eq.~(\ref{equation5}) correspond to the intensities of the signal and the second-harmonic (SH) beams, respectively \cite{kumar2020large,wang2024large}. Thus, the Hamiltonian can be rewritten as:

\begin{equation}
H = - \beta \sum_{\mu = 1}^{K} I_{\text{signal}}^{\scriptscriptstyle(\mu)} - \gamma \sum_{\mu = 1}^{K} I_{\text{SH}}^{\scriptscriptstyle(\mu)},
    \label{equation6}
\end{equation}
where \(I_{\text{signal}}^{\scriptscriptstyle(\mu)}\) and \(I_{\text{SH}}^{\scriptscriptstyle(\mu)}\) represent the intensities of the signal and SH outputs, respectively. \bigskip

Fig.~\ref{fig:OHNN-Drawing} is the workflow for implementing the NOHNN. \ref{fig:OHNN-Drawing}(a) shows some example stored patterns that generate the weights. In \ref{fig:OHNN-Drawing}(b), the neuron configuration is transformed with each stored pattern using a gauge transformation. In \ref{fig:OHNN-Drawing}(c), the transformed configurations are projected onto a spatial light modulator (SLM) sequentially while the signal and SH intensities are measured after each projection. After performing all $K$ projections, \ref{fig:OHNN-Drawing}(d) calculates the system's total energy. Next, a random neuron is altered and the steps (a)-(d) are repeated to calculate the new energy $H'$. The change $\Delta H = H'-H$ is then calculated, as shown in (e), and the updated neuron configuration is accepted if $\Delta H<0$, and rejected otherwise. The whole process is repeated for a set number of times.   

\subsection{Energy Landscape Simulation}\label{Energy Landscape Simulation}

\begin{figure}[H]
    \begin{center}
        \includegraphics[width=1.0\linewidth]{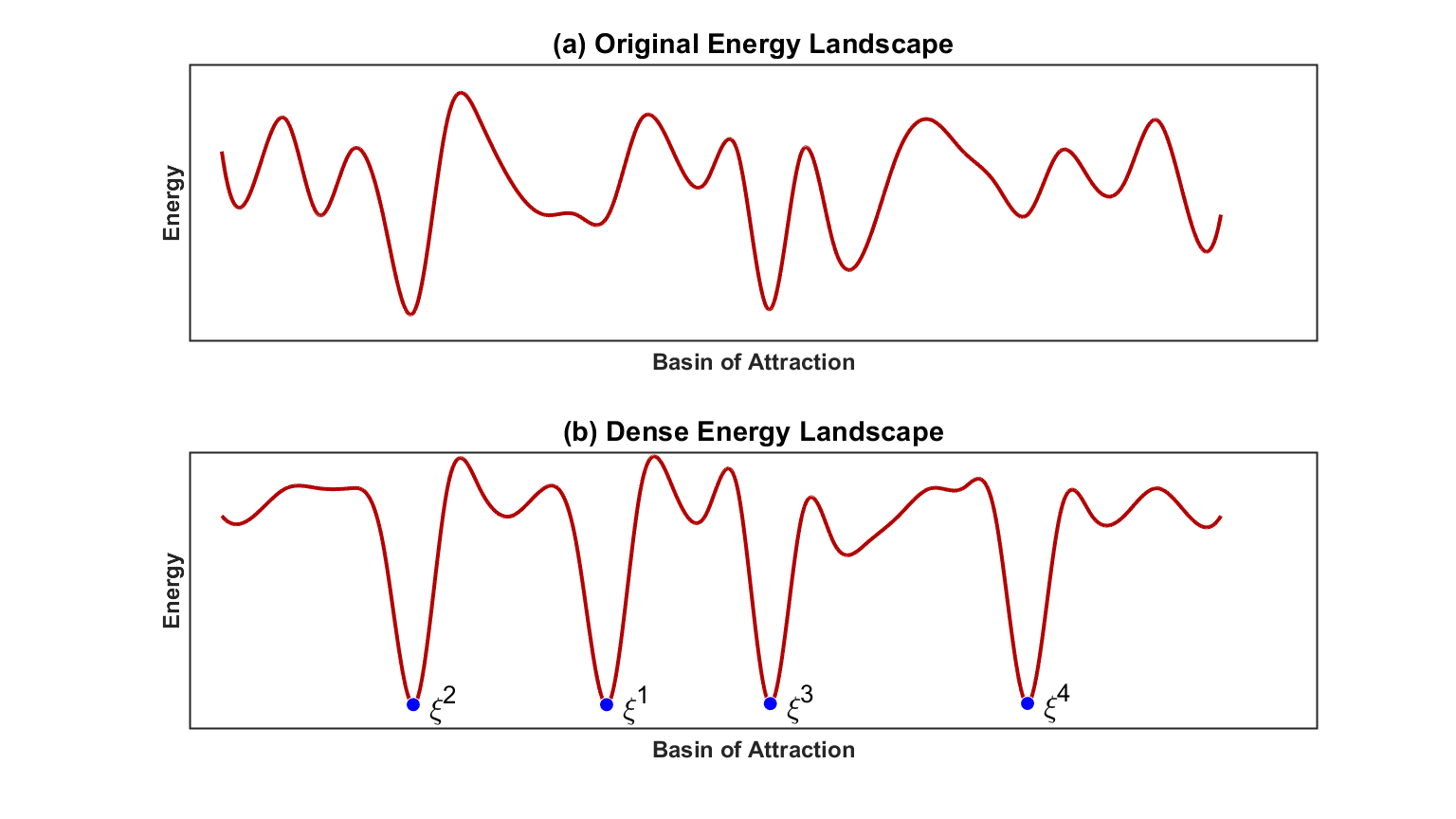}
    \end{center}
    \caption{{\bf Comparison of Energy Landscapes for traditional versus dense associative memories.} Simulated energy landscapes illustrating the basin of attraction and local minima. (a) The memory states for $n = 2$ shows the broader structure of the basins. (b) The dense energy landscape $n = 4$, corresponding to a more complex function, features steeper and sharper local minima $\boldsymbol{\xi^{\scriptscriptstyle(\mu)}}$, enabling faster convergence and the storage of more patterns/attractors with reduced cross-talk compared to traditional Hopfield networks.}
    \label{fig:energy}
\end{figure}

The enhanced memory capacity of the 4-body interaction model arises from its ability to reshape the system's energy landscape. In AM models, the energy landscape governs the stability of stored patterns (attractors) and determines how efficiently the system can retrieve them. By introducing higher-order interactions, such as the 4-body term, the model creates a more complex and structured landscape characterized by deeper and steeper local minima. For illustration, consider a small system of $N = 4$ neurons that gives $2^4$ different binary states. Two states are randomly chosen as the stored patterns and are used to compute the energy landscape from Eq.~(\ref{equation3}). As shown in Fig.~\ref{fig:energy}, the original ($n=2$) and dense ($n=4$) energy landscapes are plotted across all states to show the stable regions of our system. The 4-body energy landscape exhibits significantly steeper gradients compared to the 2-body model, resulting in sharper minima. These enhance the stability of stored patterns, making them more robust to noise while enabling faster convergence during retrieval. \bigskip

As such, a significant advantage of the 4-body model is the reduced cross-talk among stored patterns. This is because in traditional models, overlapping minima can cause the system to converge to incorrect patterns during the retrieval. The 4-body interaction mitigates this issue by increasing the energy barriers between attractors, thereby effectively isolating the basins of attraction and minimizing the cross-talk \cite{krotov2016dense}. This reduction in cross-talk is further amplified for orthogonal patterns, such as those derived from Hadamard matrices, which are square matrices with mutually orthogonal rows and columns \cite{hedayat1978hadamard}. \bigskip

\section{Experimental setup}

\begin{figure}[H]
   \begin{center}
       \includegraphics[width=\linewidth]{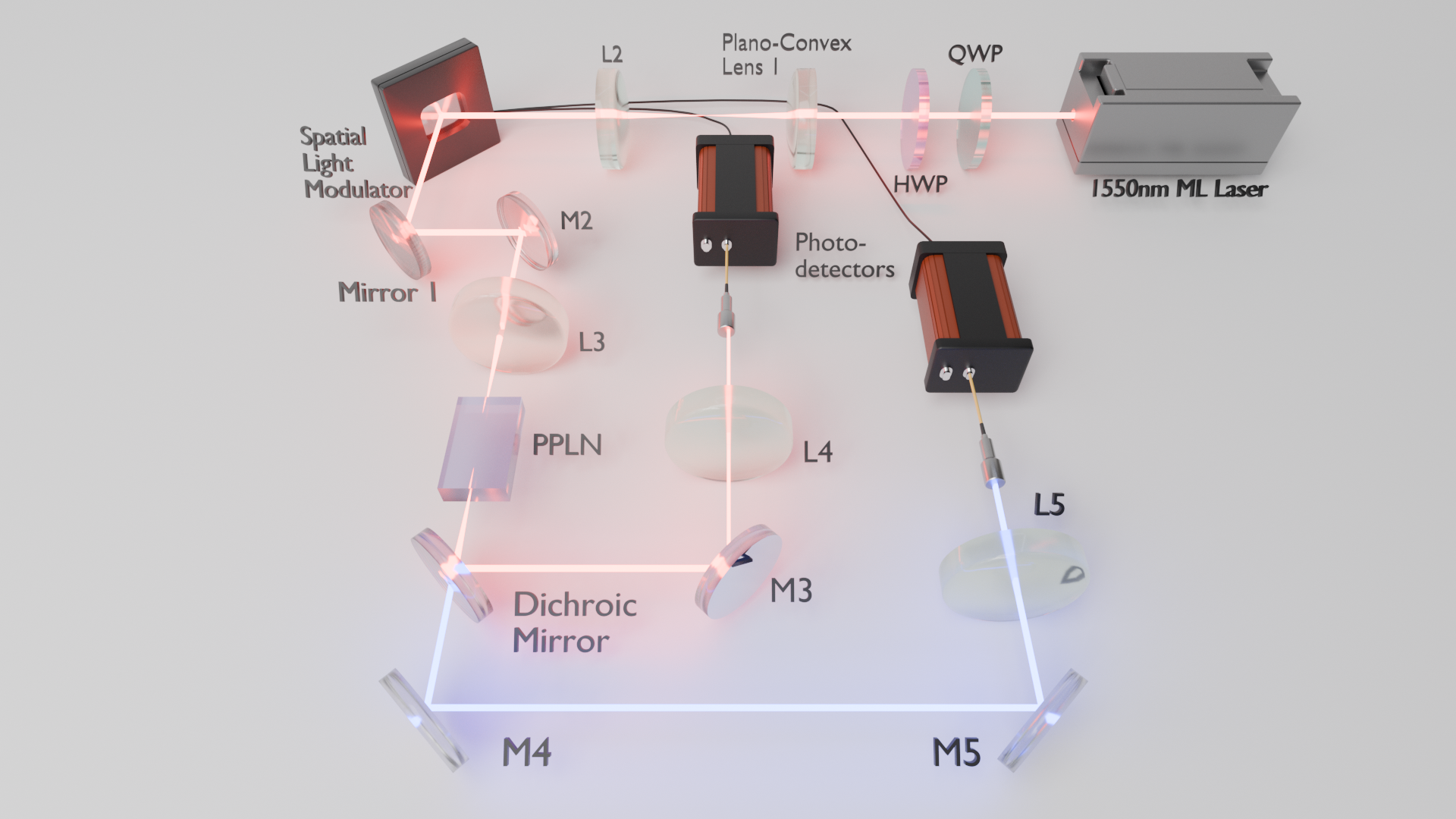} 
  
    \caption{{\bf Diagram of the experimental setup used to implement the NOHNN.} Linearly polarized light is incident on the SLM, which modulates the beam into multiple diffraction orders. A pinhole is used to isolate the first diffraction order, which is focused onto the PPLN crystal to get SH light. The signal and SH light are then coupled into single-mode fibers and measured using photodiodes. Abbreviations: PPLN – Periodically Poled Lithium Niobate, HWP – Half-Wave Plate, QWP – Quarter-Wave Plate, PBS – Polarizing Beam Splitter, SLM – Spatial Light Modulator, M – Mirror, L – Lens, D – Detector.}
    \label{fig_exp}

   \end{center}
\end{figure}

The experimental setup, depicted in Fig. \ref{fig_exp}, consists of a femtosecond mode-locked fiber laser (Calmar FPL-03CFF) emitting spectrally-broadband laser with a 50\,MHz repetition rate. The pulses are sent through a wavelength division multiplexer (WDM, 1.33\,nm bandwidth) to filter the wavelength to $\lambda_{\text{signal}} = 1550.9\,\text{nm}$. The filtered pulses, with about $4\,$ps full-width at half-maximum (FWHM), are amplified by an erbium-doped fiber amplifier (EDFA) to an average power of about 95\,mW. The laser beam passes through a combination of a quarter-wave plate (QWP) and a half-wave plate (HWP) to control polarization and modulate intensity \cite{kumar2020large,kumar2023observation}. Approximately $5\%$ of the beam is split off through a beam splitter (omitted from the figure for brevity) to monitor power fluctuations throughout experimental runs. Afterwards, a PBS directs linearly polarized light toward the nonlinear-optical setup, while the reflected arm serves as feedback for a polarization stabilizer. The transmitted portion is then expanded using a 4f system with $2 \times$ magnification to a FWHM of 3.72\,mm, before directed onto the SLM (Santec SLM-100, 10.4\,$\mu$m pixel pitch, about 300 ms response rate). The SLM displays a 10$\times$10 array of macro-pixels (each containing 20$\times$20 pixels) to encode the stored pattern, spin configuration, and a blazed grating to separate the diffraction orders. A gauge transformation~\cite{fang2021experimental,katidis2024robust} enables these screen projections on the SLM.

Input patterns are stored in the system through time division multiplexing (TDM) \cite{yamashita2023low,yamamoto2017time} for algorithmic simplicity and performance efficiency in our optical architecture. We align the rest of the optical setup using the first-order diffracted light to minimize edge diffraction effects. Two aligning mirrors then direct the light into another 4f system, focusing it at the Fourier plane, where the temperature-controlled PPLN crystal is placed for maximum SH efficiency. A dichroic mirror then separates the signal light from the SH light. Finally, the light in each of the reflected and transmitted arms is coupled into a single-mode fiber with a coupling efficiency at or above 70\%. The intensities are measured with photodiodes (Visible: Thorlabs PDA8A2, IR: Thorlabs PDA20CS2), which are then connected to a National Instruments Data Acquisition (DAQ) USB Device to record the voltages. The DAQ connects to a laptop running a MATLAB interface that updates the SLM in real time, to complete the feedback loop.

\section{Results}\label{Results}

The experiment is conducted using three types of patterns: orthogonal Hadamard (uncorrelated), correlated, and MNIST digit patterns \cite{6296535}. Each pattern type is carefully selected to demonstrate specific properties of the DAM. Orthogonal Hadamard patterns highlight the system's increased pattern capacity, while correlated patterns emphasize the impact of correlation in the system's performance. MNIST digit patterns showcase the DAM's robustness to noise fluctuations arising from varied correlations in real-world scenarios. All experiments are performed under the same conditions, with a masking percentage of $\delta = 0.20$, number of neurons $N=100$, and the same number of iterations per run ($T=N$) (expect for some additional results in the Appendix \ref{app:digital_emulation}). For each pattern type, the experiment is carried out with two different orders of interaction $n$. The first run exclusively involves only 2-body interaction, with $\beta = 1$ and $\gamma = 0$ in Eq.~(\ref{equation6}). The second run combines 2-body and 4-body interactions, with $\beta = 1$ and $\gamma = 200$. Here, the 4-body scaling constant is chosen to be much greater than that of the 2-body scaling constant ($\gamma\gg\beta$), to ensure that the system's dynamics and evolution are primarily driven by the SH optical nonlinear effect $I_{\text{SH}}$ rather than the signal intensity $I_{\text{signal}}$. In the following, runs using the combined 2-body and 4-body interaction Hamiltonian in Eq.~(\ref{equation6}) will be referred to as the 4-body interaction run.

\subsection{Hadamard Patterns}

Constructing stored patterns from Hadamard matrices offers a significant advantage in memory recall. The key feature of a Hadamard matrix is that its columns and rows are mutually orthogonal. This means that the inner product of any two distinct stored patterns yields a Kronecker delta, i.e., \( \boldsymbol{\xi^{\scriptscriptstyle (\mu)}} \cdot\boldsymbol{\xi^{\scriptscriptstyle (\nu)}} = N\delta_{\mu \nu}\). This results in nearly vanishing cross-correlations among the patterns. As a result, both the mean and standard deviation of the correlation are zero, thereby eliminating noise arising from the cross-talk. \bigskip

\begin{figure}[H]
   \centering
       \includegraphics[width=\linewidth]{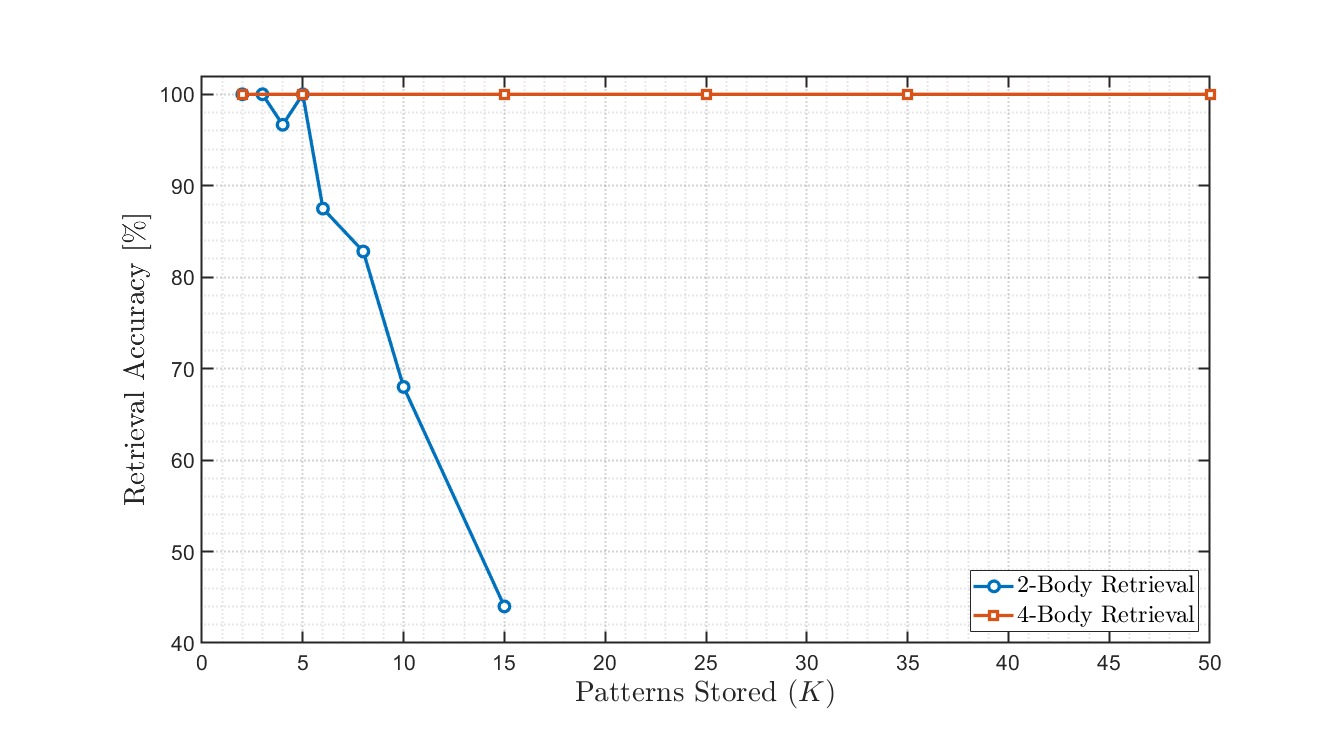} 

    \caption{{\bf Enhanced Retrieval Capacity of 4-Body vs. 2-Body on Hadamard Patterns.} Retrieval accuracy \(\eta\) of traditional AM (blue) and DAM (red) as a function of the number of patterns stored $K$, using Hadamard patterns. The plot demonstrates a minimum tenfold improvement in the critical capacity (maximum number of patterns stored), with $K_c^{\scriptscriptstyle (n = 4)} \geq 50$ for the 4-body interaction model, compared to $K_c^{\scriptscriptstyle (n = 2)} = 5$ for the traditional 2-body model.}
    \label{fig_had_optical}

\end{figure}

To calculate the retrieval accuracy $\eta$ for a system with stored Hadamard patterns, we use the similarity threshold $\tau = 0.60$ to define sufficient distance between the retrieved pattern and the other stored patterns. Figure~\ref{fig_had_optical} shows that by increasing the order of interaction of the system, the critical pattern capacity \(K_{c}\) increases by at least a factor of $10$, with $K_c^{\scriptscriptstyle (n = 4)} / K_c^{\scriptscriptstyle (n = 2)} \geq 10$. While this increase factor could potentially be much larger \cite{krotov2016dense}, verifying this experimentally would require testing the system with a significantly higher number of stored patterns. However, the limited refresh rate and response time of the SLM used in this setup make such experiments time-prohibitive. This challenge can be addressed by utilizing faster, modern SLMs~\cite{ye2021high} or even digital micromirror devices (DMDs)~\cite{mitchell2016high}, which are now commercially available and would enable more extensive testing and improved capacity. \bigskip

To further validate the above, we construct a second setup using a similar energy model without the optical SH nonlinearity. By utilizing a faster visible SLM (HOLOEYE, about 150\,ms response rate), we overcame the original time constraints and repeated the testing. The results are given in Appendix \ref{app:digital_emulation}, which reveal a critical capacity ratio of \(K_c^{\scriptscriptstyle (n = 4)} / K_c^{\scriptscriptstyle (n = 2)} \geq 28.8\). These results demonstrate the increased storage capacity due to the 4-body interaction. 

\subsection{Correlated Random Patterns}
\begin{figure}[H]

    \begin{subfigure}[b]{0.49\linewidth}
      \includegraphics[width=\linewidth]{"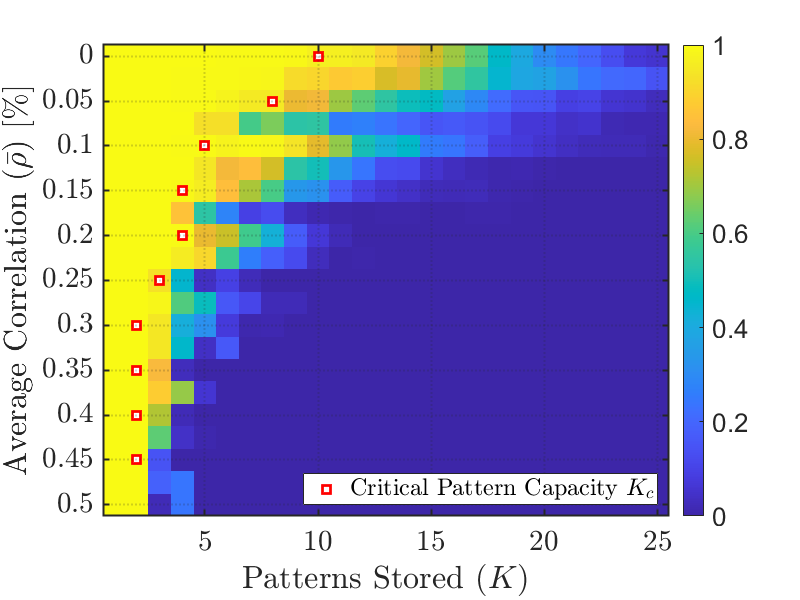"} 
      \label{fig:2_body_phase_transition}
   \end{subfigure}
   \hfill
   \begin{subfigure}[b]{0.49\linewidth}
      \includegraphics[width=\linewidth]{"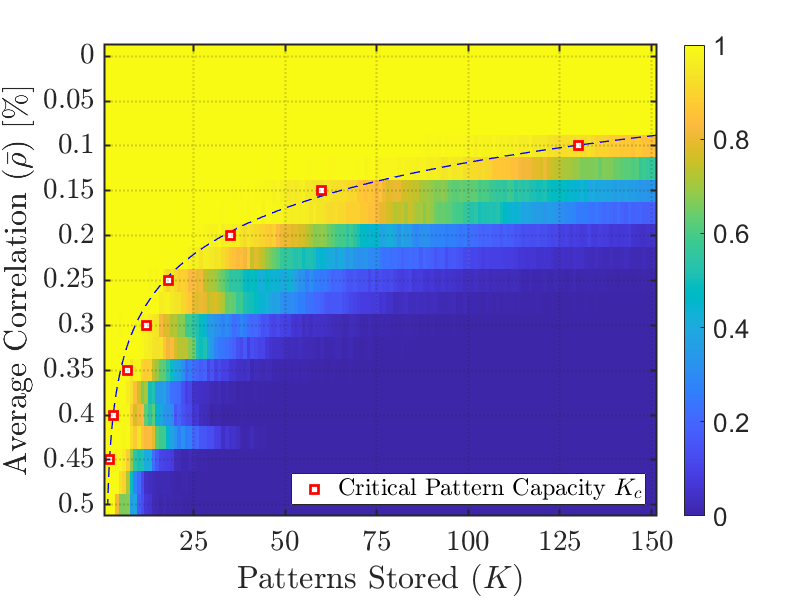"} 
      \label{fig:phase_transition_curve}
   \end{subfigure}
   
   \caption{{\bf Pattern storage capacity as a function of average correlation for 2-body and 4-body interactions.} Two simulation color maps for the number of patterns stored \(K\) as a function of average correlation \(\bar{\rho}\), demonstrating a distinct phase transition curve. Left color map is formed using a 2-body interaction Hamiltonian, whereas the right one is formed using a 4-body term only. The color bar represents the retrieval accuracy of the system, averaged over 50 trials per \((K, \bar{\rho})\) parameter set. Experimental data points (red open squares), representing the critical capacity \(K_{c}\) for each correlation value, are overlaid to compare with the numerical simulation. Gray dashed line is the fitting curve according to Eq.~(\ref{fitting}).}
   \label{fig_phase_transition}
\end{figure}

Next, the system is tested on correlated patterns. In each trial, a set of $K$ random patterns is generated with a fixed average correlation \( \bar{\rho} \); see Appendix~\ref{matlab} for definition and details. Then, the DAM is simulated numerically for $T=100$ iterations following the workflow in Fig.~\ref{fig:OHNN-Drawing}. The simulation is performed over 50 trials for each $(K,\bar{\rho})$ configuration, and the resulting retrieval accuracy \( \eta \) is computed as the average over all trials. The mean retrieval accuracies are plotted as heat maps, as shown in Fig.~\ref{fig_phase_transition}. The left map corresponds to a 2-body Hamiltonian (\(\beta = 1\) and \(\gamma = 0\) in Eq.~(\ref{equation5})), while the right one corresponds to a 4-body Hamiltonian (\(\beta = 0\) and \(\gamma = 1\)). Finally, the experimental data points representing the critical pattern capacity \( (K = K_{c}) \) are plotted against their respective average correlation. \bigskip

When the fluctuations in the cross-correlation between patterns ($\rho_{\mu \nu}$) are small, the critical capacity of the system $K_c$ is predominantly determined by $\bar{\rho}$. As shown in Fig.~\ref{fig_phase_transition}, $K_c$ decreases exponentially with $\bar{\rho}$ for the 4-body system. To quantify it, the experimental data is fitted as:
\begin{equation}
    K^{(n=4)}_c = (C_1-1)e^{- C_2 \bar{\rho}} + 1,
    \label{fitting}
\end{equation}
where $C_1$ is the 4-body critical pattern capacity for $\bar{\rho} = 0$, and $C_2$ is the decay factor. The fitting yields $C_1 = 508.4 \pm 74.10$ and $C_2 = 13.82 \pm 1.210$ with 95\% confidence. For $\bar{\rho} = 0$, the 2-body system has a critical pattern capacity of 10. Therefore, the 4-body interaction lifts the analytical bound for the critical capacity by about 50 times. This advantage persists across a range of $\bar{\rho}$, with the 4-body system maintaining a far higher critical capacity. Additionally, it exhibits enhanced sensitivity in $\eta$ to variations in $\bar{\rho}$, highlighting its robustness and scalability in noisy regimes.

\subsection{MNIST Handwritten Digit Patterns}

For benchmark against well-established patterns, the system is further tested with MNIST handwritten digit patterns. In each run, a set of $K$ patterns is constructed by selecting a digit from each class (0--9). For $K > 10$, additional variations are added by cycling through the classes again. To ensure the number of neurons is $N=100$, each MNIST pattern is digitally resized from a $28\times28$ to a $10\times10$ using MATLAB's \texttt{imresize} function with bicubic interpolation. These chosen patterns exhibit high average correlation \(\bar{\rho} \approx 0.22\), as well as visual intuitiveness that allows for straightforward comparisons between replicas and the ground truth. Furthermore, patterns within the same digit class can have \(\rho_{\mu \nu}\) as high as 0.7; see Appendix \ref{Retrieval at Critical Capacity} for more details. These spikes increase the standard deviation of the correlation and test the system's robustness in handling highly overlapping patterns. \bigskip

\begin{figure}[H]
   \centering
       \includegraphics[width=\linewidth]{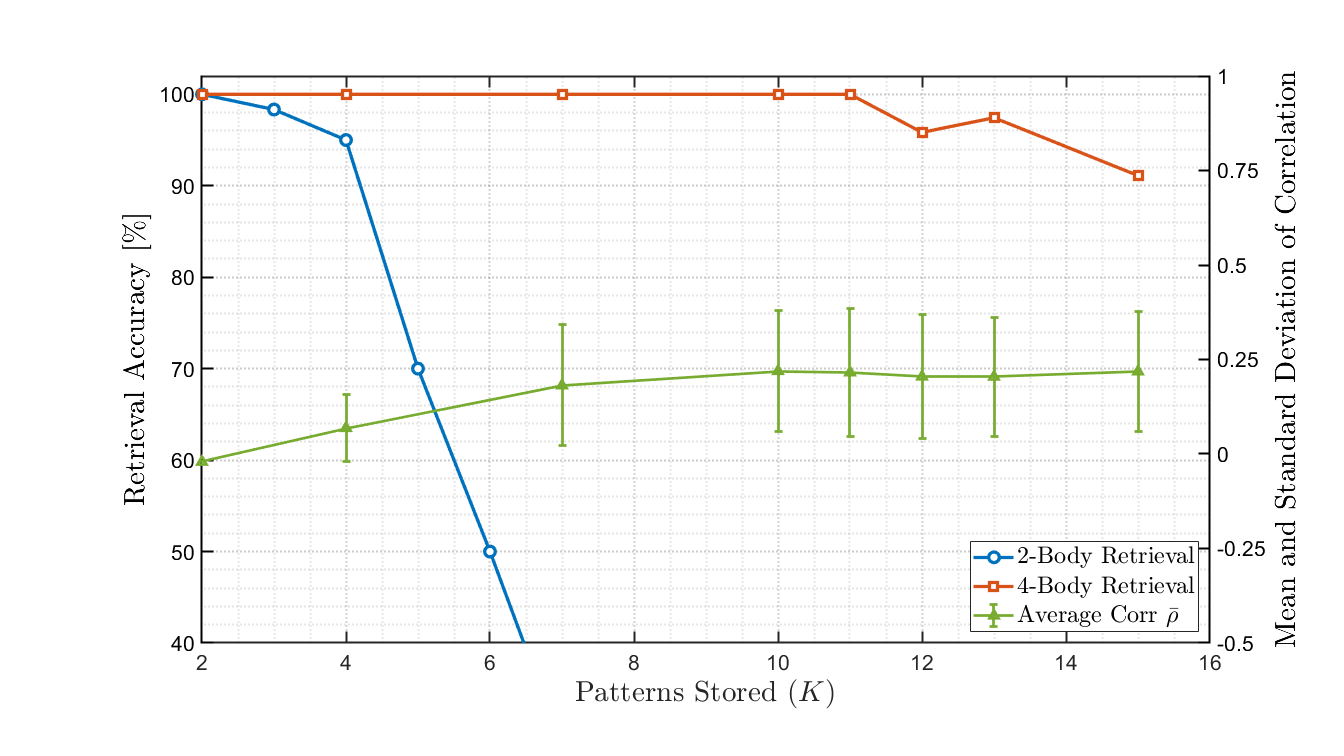} 

    \caption{{\bf Retrieval accuracy of 4-Body vs. 2-Body on MNIST Patterns.} Retrieval accuracy $\eta$ (left y-axis) of traditional AM (blue) and DAM (red) as a function of $K$, using MNIST digit patterns. The plot demonstrates a $5.5$ times improvement in the critical capacity, with $K_c^{\scriptscriptstyle (n = 4)} = 11$ for the 4-body interaction model, compared to $K_c^{\scriptscriptstyle (n = 2)} = 2$ for the traditional 2-body model. The average correlation \( \bar{\rho} \) and standard deviation (right y-axis) are also shown, highlighting the 4-body model's robustness to random noise fluctuations.}
    \label{fig_MNIST_optical}

\end{figure}

For MNIST digits, a similarity threshold \(\tau = 0.20\) is used to differentiate the converged pattern from the other stored patterns while accounting for cross-correlations. Figure \ref{fig_MNIST_optical} shows \(\eta\) as a function of the number of \(K\) for both the 2-body and 4-body energy models. The results reveal a critical capacity ratio \(K_c^{\scriptscriptstyle (n = 4)} / K_c^{\scriptscriptstyle (n = 2)} = 5.5\), demonstrating again the superior scalability of the 4-body model. Additionally, the large average correlation \(\bar{\rho}\) and its standard deviation (right y-axis) in those patterns showcase the system's robustness to noise fluctuations caused by pattern correlations. \bigskip

\begin{figure}[tbph]
    \centering
    \includegraphics[width=0.75\linewidth]{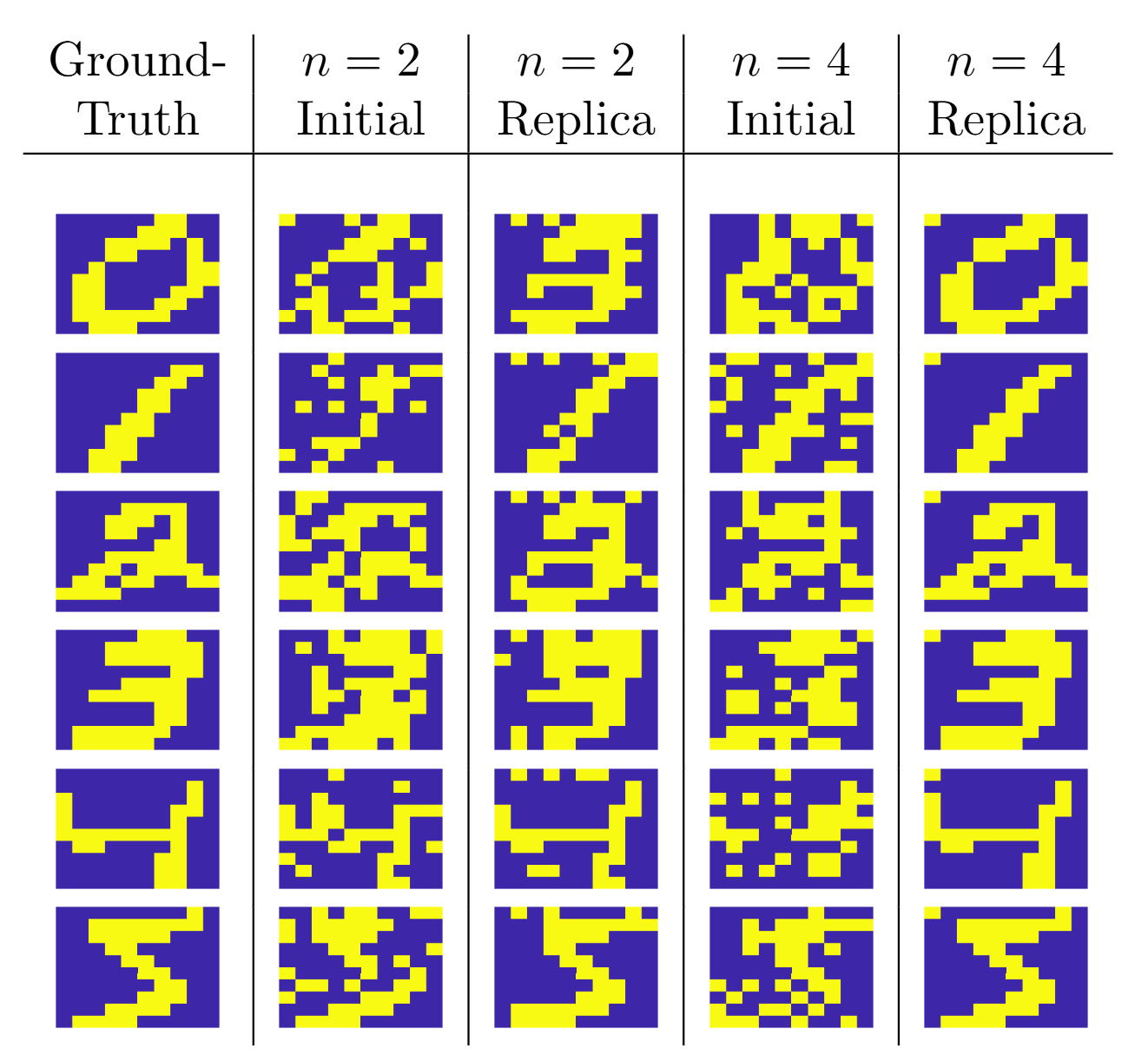}
    
\caption{{\bf Retrieval of MNIST Digits in 2-Body vs 4-Body.} Comparison of pattern retrieval performance for $K = 6$ MNIST digits using 2-body and 4-body interactions. Each row represents a digit (0-5) where: Column 1 shows the ground-truth pattern ($\boldsymbol{\xi^{\scriptscriptstyle(\mu)}}$); Columns 2-3 show the initial corrupted pattern $\boldsymbol{\xi_{\text{mask}}}$ and final retrieved configuration (replica) for the 2-body model; Columns 4-5 show the corresponding patterns for the 4-body model.}
    \label{fig:stored_patterns}
\end{figure}

As an example, Fig.~\ref{fig:stored_patterns} compares the pattern retrieval performance in a single run for \(K = 6\) MNIST digits using the two energy models when the initial patterns are randomly masked by 20\%. As shown, the replicas produced by the 2-body model are noisy, where that for digit~``0'' could be mistaken as ``3'', and digits ``2'' and ``3'' are unrecognizable. The retrieval accuracy, calculated as the number of correct retrievals (following the retrieval conditions from Section~\ref{Retrieval accuracy}) over all attempts, is only \(\eta = 3/6 = 0.50\). In contrast, the 4-body model performs much better, where the noise is negligible and the retrieval accuracy is \(\eta = 1.0\). This demonstrates the 4-body system’s enhanced robustness, achieving cleaner and more accurate reconstructions. More examples of MNIST retrieval with the 4-body energy model can be found in Appendix \ref{MNIST Digits with Varying Pattern Capacity}.

\section{Conclusion}\label{sec13}

We have experimentally demonstrated, for the first time, a nonlinear optical system for realizing DAMs with both 2-body and 4-body interactions. For uncorrelated patterns, it achieves over tenfold improvement in the critical pattern capacity over those with only 2-body interaction. For correlated patterns, the improvement can reach $50$ times,  depending on the average correlation among them. Testing on MNIST handwritten digit patterns, it achieves a $5.5$ times improvement with cleaner retrieved patterns, which indicates its scalability and robustness for practical applications.
These findings underscore the potential of nonlinear-optical DAMs for real-world applications in complex optimization, computer vision, and graph network analysis. \bigskip

In the future, this system can be improved for real time applications by employing high-speed SLMs, such as those running at GHz capable of completing trials on the order of milliseconds \cite{smolyaninov2019programmable,liu2017focusing}. Also, while the above demonstrations have used time division multiplexing, our technique can be applied to alternative implementations such as focal plane division (FPD) \cite{veraldi2025fully}, wavelength division multiplexing (WDM) \cite{luo2023wavelength}, and space division multiplexing (SDM) \cite{sakabe2023spatial}. To further improve the storage capacity or incorporate quantum effects to accelerate the pattern retrieval  \cite{fiorelli2019quantum}, one may implement an exponential energy function \cite{ramsauer2020hopfield} by using an optical parametric amplifier \cite{kumar2020large,kumar2023observation}. Based on our results and with those further development, non-Von Neumann, neuro-inspired photonic computing \cite{brunner2025roadmap} can be developed as an energy-efficient alternative for high-capacity memory storage and a mean to simplify deep learning neural architectures for advanced AI applications \cite{jordan2015machine}. \bigskip

%Also, integrated nanophotonic recurrent Ising sampler (INPRIS) chips, which also support GHz clock rates, can be improved by to storability  enabling real-time data processing \cite{prabhu2020accelerating}. 

%An alternative variant of spatial-photonic Ising machines can be implemented which increases the degrees of freedom from the coupling matrix rather than the energy complexity, ultimately improving performance \cite{wang2025efficient}. 

\backmatter

\bmhead{Data availability}

The data collected for this study is publicly available in the GitHub repository at https://github.com/kmusa11/OHNN-CDAM.git \cite{Musa_Correlated_Dense_Associative_2025}. 

\bmhead{Code availability}

The MATLAB scripts and results used are publicly available in the GitHub repository at https://github.com/kmusa11/OHNN-CDAM.git \cite{Musa_Correlated_Dense_Associative_2025}. 

\bmhead{Acknowledgments:}
We thank Dr. Chunlei Qu and Mr. Zhaotong Li for inspirational discussions and collaborations.  

\bmhead{Author Contributions}
KM conducted the experiment with help from SK and MK under the supervision of YH. KM analyzed data and performed simulation. All contributed to the inception and execution of this project, and KM, YH, and SK authored the manuscript.   

\section*{Declarations}

\bmhead{Funding}

This work is supported in part by the ACC-New Jersey under Contract No. W15QKN-18-D-0040.

\bmhead{Conflict of interest}

The authors declare no conflict of interest.

%\bmhead{Author contribution}

\newpage

\begin{appendices}

\section{NOHNN Theoretical Analysis} \label{theoretical}

In this section, we analyze the behavior of the electric field incident on the SLM, and establish a connection between the measured intensities and the theoretical framework of both the traditional and dense HNN. To derive the NOHNN Hamiltonian, we model the incident electric field as a Gaussian signal beam with wavelength $\lambda_{\text{s}}$:

\begin{equation}
    E_{\text{signal}}(x,y) = E_0 e^{-\frac{x^2 + y^2}{w_0^2}}
\end{equation}

where $w_0$ is the beam waist and $E_0$ is the peak amplitude. The SLM then imposes a spatially varying phase modulation $\varphi_{nm}$ on the incident beam according to the gauge transformation:

\begin{equation}
    \varphi_{nm} = \sigma_{nm}\frac{\pi}{2} + (-1)^{n+m+1} \arccos(\xi_{nm}),
\end{equation}

where, $n = 1,2,\dots,{\tilde{N}}$, $m = 1,2,\dots,{\tilde{M}}$, $N = {\tilde{N}} \times {\tilde{M}}$, $\sigma_{nm} \in [-1, +1]$ represents the neuron state at pixel $(n,m)$, and $\xi_{nm}$ denotes the input pattern value at the corresponding pixel location. The modulated electric field after the SLM can be written as:

\begin{equation}
    E_{\text{signal}}'(x,y) = \sum_{n=1}^{\tilde{N}} \sum_{m =1}^{\tilde{M}} g_{nm} e^{i\varphi_{nm}} \frac{1}{d^2} \, \Pi\left(\frac{x - x_m}{d}\right)\Pi\left(\frac{y - y_n}{d}\right),
\end{equation}

with $g_{nm} \coloneqq E_0 e^{- \left( \frac{x_m^2 + y_n^2}{w_0^2}\right)}$ defined as the gaussian weight of each pixel, $\Pi(\cdot)$ represents the rectangular function, $x_m \coloneqq (m-\frac{1}{2})d$ and \(y_n \coloneqq (n-\frac{1}{2})d\) denote the pixel center positions, and $d$ is the pixel pitch. Upon propagation through a Fourier lens of focal length \(f\), the field at the focal plane is given by its Fourier transform:

\begin{align}
    &E_{\text{signal}}''(u,v) = \sum_{n=1}^{\tilde{N}} \sum_{m =1}^{\tilde{M}} g_{nm} e^{i\varphi_{nm}} \frac{1}{d^2} \, \mathscr{F} \left\{ \Pi\left(\frac{x - x_m}{d}\right)\Pi\left(\frac{y - y_n}{d}\right) \right\}(u,v) \nonumber \\
    &= \sum_{n=1}^{\tilde{N}} \sum_{m =1}^{\tilde{M}} g_{nm} e^{i\varphi_{nm}} \frac{1}{d^2} \iint_{-\infty}^{\infty} \Pi\left(\frac{x - x_m}{d}\right)\Pi\left(\frac{y - y_n}{d}\right) \, e^{-\frac{2 \pi i}{\lambda_s f} (xu + yv)}   \, dx \, dy \nonumber \\
    &= \sum_{n=1}^{\tilde{N}} \sum_{m =1}^{\tilde{M}} i g_{nm} \xi_{nm} \sigma_{nm}  \Lambda_{nm} \operatorname{sinc}(\frac{\pi ud}{\lambda_s f}) \operatorname{sinc}(\frac{\pi vd}{\lambda_s f}),
\end{align}

where $\Lambda_{nm} \coloneqq e^{-\frac{2 \pi i}{\lambda_s f} (x_mu + y_nv)}$. The phase term simplifies to $e^{i \varphi_{nm}} = i \, \xi_{nm} \, \sigma_{nm} $ due to the binary nature of the variables. \bigskip

Since only the near axis light gets coupled into the fiber for detection, we can assume the value $ud << \lambda_s f$, which can simplify $\operatorname{sinc}(\frac{\pi ud}{\lambda_s f}) \approx 1$. Another approximation can be made to the weights $g_{nm}$, stating the area of the gaussian beam is much larger than the region of interest where we display our patterns $\pi w_0^2 >> d^2N$. This approximation allows us to assume uniform amplitude across all pixels $g_{nm} \approx E_0$, only if the display region ($d^2N$) is placed at the center of the gaussian beam. Finally, we perform an index transformation $j = m + (n-1){\tilde{N}}$ for simplicity, so the final electric field at the focal point (where the nonlinear crystal is placed) becomes: 

\begin{equation}
    E_{\text{signal}}''(u,v,z) = i E_0 \sum_{j=1}^{N} \xi_{j} \sigma_{j} \Lambda_{j} e^{ik_sz},
    \label{eqnA5}
\end{equation}

where \(k_s = 2\pi/\lambda_s\) is the signal wave-vector. This field interacts with a periodically poled Lithium Niobate (PPLN) crystal, generating second-harmonic light (\(E_{\mathrm{SH}}\)) via the coupled nonlinear equations:

\begin{align}
    &2ik_s \frac{\partial E_{\text{signal}}}{\partial z} + \left( \frac{\partial^2}{\partial x^2} + \frac{\partial^2}{\partial y^2} \right) E_{\text{signal}} = -2\frac{\omega_s^2}{c^2}\chi^{(2)}E_{\text{signal}}^*E_{\text{SH}} e^{i\Delta\kappa z}, \\
    &2ik_h \frac{\partial E_{\text{SH}}}{\partial z} + \left( \frac{\partial^2}{\partial x^2} + \frac{\partial^2}{\partial y^2} \right) E_{\text{SH}} = -\frac{\omega_h^2}{c^2}\chi^{(2)}E_{\text{signal}}^2 e^{-i\Delta\kappa z},
\end{align}

where \(k_h= 2\pi n_c/\lambda_h\) is the SH wave-vector in the medium, $n_c$ is the index of refraction of the crystal, \(\omega_h = 2 \pi c/\lambda_h \) is the SH angular frequency, and \(\Delta\kappa = 2k_s - k_h\) is the phase mismatch. By assuming phase matching condition $\Delta \kappa = 0$ and undepleted pump approximation, the SH electric field at the output of the crystal (assuming the center of the crystal is at $z=0$) can be solved analytically as:

\begin{equation}
    E_{\text{SH}}(u,v,L/2) = i \frac{\omega_{h} \chi^{(2)}L}{2c^2k_h} E_{\text{signal}}^2(u,v,-L/2)
    \label{eqnA8}
\end{equation}

Plugging Eq.~(\ref{eqnA5}) to Eq.~(\ref{eqnA8}) we get:

\begin{equation}
    E_{\text{SH}}(u,v,L/2) = - \frac{E_0^2 \omega_{h} \chi^{(2)}L}{2c^2k_h} \sum_{j,k=1}^{N}  \xi_{j} \xi_{k} \sigma_{j} \sigma_{k}  \Lambda_{j} \Lambda_{k} e^{-ik_sL},
    \label{eqnA9}
\end{equation}

Finally, the signal and SH intensities can be related to their respective fields by $I\propto |E|^2$:
\vspace*{-5mm}

\begin{align}
    I_{\text{signal}} &\propto \sum_{i,j=1}^{N} \xi_{i} \xi_{j} \, \sigma_{i} \sigma_{j}  \label{eqnA10} \\ 
    I_{\text{SH}} &\propto \sum_{i,j,k,l=1}^{N} \xi_{i} \xi_{j} \xi_{k} \xi_{l} \, \sigma_{i} \sigma_{j} \sigma_{k} \sigma_{l} \label{eqnA11}
\end{align}

The rest of the coefficients in front of the intensity expression are just constants so we can disregard them as they play no significant role. The exponential terms, as well as $\Lambda$, all cancel out ($\Lambda\Lambda^* = 1$).  These expressions map directly to the 2-body and 4-body interaction terms of the HNN energy model, demonstrating the optical realization of NOHNN dynamics.

\newpage
\section{Correlation in Patterns} \label{Correlation in Patterns}

The construction of patterns significantly influences the memory capacity of the system. This influence arises from the similarity between the patterns, which we formally define as correlation. Intuitively, if two patterns are highly correlated, they encode similar information, making it difficult to distinguish one from the other. This redundancy reduces the effective memory capacity of the system. Therefore, using uncorrelated patterns is the ideal case for achieving the highest memory capacity. The correlation matrix \(\boldsymbol \rho\) between the stored patterns \(\boldsymbol \xi\) is defined using the Pearson Correlation Coefficient (PCC) \cite{wu2010correlation} as follows:

\begin{equation}
   \rho_{\mu \nu} = \frac{\frac{1}{N} \left( \boldsymbol{\xi^{\scriptscriptstyle(\mu)}} \cdot \xi^{\scriptscriptstyle(\nu)} \right) - \bar{\xi}^{\scriptscriptstyle(\mu)} \bar{\xi}^{\scriptscriptstyle(\nu)}}{\sqrt{\left( 1 - (\bar{\xi}^{\scriptscriptstyle(\mu)})^2 \right) \cdot \left( 1 - (\bar{\xi}^{\scriptscriptstyle(\nu)})^2 \right)}},
\end{equation}

where \(\boldsymbol{\xi^{\scriptscriptstyle(\mu)}}\) and \(\xi^{\scriptscriptstyle(\nu)}\) represent the \(\mu\)-th and \(\nu\)-th patterns, respectively, \(N\) is the dimensionality of the patterns, and \(\bar{\xi}^{\scriptscriptstyle(\mu)} = \frac{1}{N} \sum_{i=1}^{N} \xi_i^{\scriptscriptstyle(\mu)}\) denotes the mean of the \(\mu\)-th pattern. The numerator describes the covariance between the two patterns \(\boldsymbol{\xi^{\scriptscriptstyle(\mu)}}\) and \(\xi^{\scriptscriptstyle(\nu)}\), while the denominator normalizes the correlation by the product of their standard deviations, ensuring that \(\rho_{\mu \nu}\) lies in the range \([-1, 1]\). \bigskip

Perfectly uncorrelated patterns can be generated using the orthogonality of Hadamard matrices, which are square matrices with mutually orthogonal rows and columns. For a Hadamard matrix of size \(N\), its columns can serve as a set of uncorrelated patterns \(\boldsymbol \xi\), allowing the system to store up to \(K = N-1\) patterns. However, the construction of Hadamard matrices is constrained by specific sizes. For instance, using the \textit{Paley Type I} construction, a Hadamard matrix of size \(N = q + 1\) can be constructed when \(q\) is a prime number satisfying \(q \equiv 3 \mod 4\) \cite{sapm1933121311,jones2017paleypaleygraphs,Musa_Correlated_Dense_Associative_2025}. \bigskip

While uncorrelated patterns are ideal for theoretical analysis, practical systems often involve patterns with varying degrees of correlation. This is because real-world data, such as images or signals, often exhibits inherent correlations due to underlying structure or dependencies. To better reflect these scenarios, we consider patterns that exhibit controlled correlations. One approach to generating such patterns involves sampling vectors from a random distribution of \(\{-1, 1\}\) and iteratively adjusting them until the average correlation matches a target value within a specified tolerance. Although this method is straightforward, more efficient and rigorous techniques exist \cite{Caprara2014GenerationOA}. An example of such an algorithm is provided in Listing~\ref{lst:CorrPatterns}. \bigskip

Since realistic patterns exhibit varying degrees of correlation with one another, we quantify the overall correlation of a set of patterns using the \textit{average correlation} \(\overline{\rho}\). This metric provides a single value to characterize the similarity between patterns and is computed by summing all off-diagonal elements of the correlation matrix \(\boldsymbol \rho\) and dividing by the total number of off-diagonal elements, \(K(K-1)\):

\begin{equation}
    \overline{\rho} = \frac{1}{K(K-1)} \sum_{\mu \neq \nu}^{K} \rho_{\mu \nu}.
\end{equation}

A lower \(\overline{\rho}\) indicates less redundancy and greater independence among patterns, while a higher \(\overline{\rho}\) suggests greater similarity and overlap between patterns.

\section{MATLAB Code for Generating Correlated Patterns}
\label{matlab}

This MATLAB script generates correlated patterns using a Gaussian noise-based approach following a method found in the forum \cite{mathoverflow_bernoulli}:

\begin{lstlisting}[style=matlabstyle, caption={MATLAB code for generating correlated patterns.}, label={lst:CorrPatterns}]

P = 20; % Number of patterns
N = 100; % Number of pixels
corr_initial = 0.30; % Target correlation between patterns

while true
    d = eye(P) + corr_initial * (ones(P, P) - eye(P));
    
    % Augment the matrix D with mean values
    g = [d, zeros(P, 1); zeros(1, P), 1];
    
    % Apply the transformation f(x) = sin(pi * x / 2) 
    h = sin(pi * g / 2);
    
    % Generate Gaussian random variables with covariance matrix h
    A = mvnrnd(zeros(1, P+1), h, N);
    
    % Binarize the Gaussian samples using the sign function
    b = sign(A);
    
    % Construct the patterns by multiplying (P+1)-th column
    c = b(:, 1:P) .* b(:, P+1);
    
    % Calculate the correlation matrix of the generated patterns
    corr = calculate_correlation_matrix(c);
    
    % Compute the average off-diagonal correlation
    avg_corr = sum(sum(corr - eye(P))) / (P^2 - P);
    
    if abs(avg_corr - corr_initial) < 1e-4
        break; % Exit if target correlation is achieved
    end
end
\end{lstlisting}

\section{MNIST Digits with Varying Pattern Capacity $K$} \label{MNIST Digits with Varying Pattern Capacity}

In this section, we analyze the pattern retrieval performance in the 4-body energy model for MNIST digit patterns under different pattern capacities $K$, specifically at the critical capacity $K = K_c$ and above it ($K > K_c$). Fig.~\ref{fig_MNIST_hybrid} illustrates the retrieval outcomes, with an average correlation value of $\bar{\rho} \approx 0.22$ across the dataset. 

\subsection{Retrieval at Critical Capacity ($K = K_c$)}\label{Retrieval at Critical Capacity}

\begin{figure}[H]
    \centering
    \includegraphics[width=0.75\linewidth]{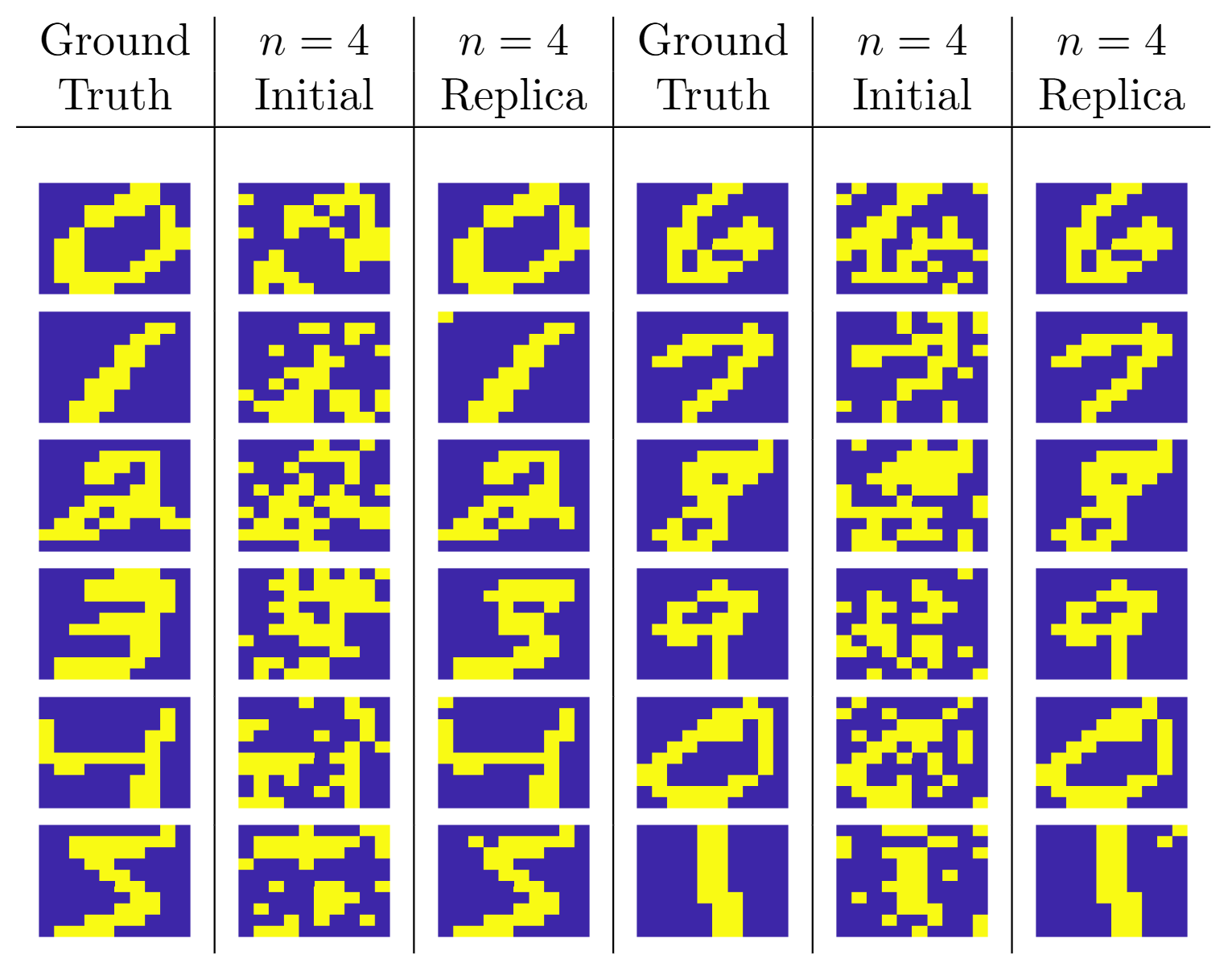}
    
\caption{{\bf Retrieval at Critical Capacity ($K = K_c$).} Pattern retrieval performance at critical pattern capacity $K_c^{\scriptscriptstyle(n=4)} = 12$ for MNIST digits using the 4-body model. This trial shows failed retrieval for digit 3 which is mixed with digits 5 and 8 based on the overlap matrix $\boldsymbol m$.}
    \label{fig:stored_patterns_2}
\end{figure}

At the critical capacity $(K = K_c)$, we observe the onset of retrieval accuracy degradation, characterized by $\eta < 1$. This phenomenon is depicted in Fig.~\ref{fig:stored_patterns_2}. Most replicas successfully retrieve the masked patterns, with the exception of digit $3$. The replica corresponding to digit $3$ exhibits partial retrieval, mixing with digits $5$ and $8$. This behavior can be interpreted as the system becoming trapped in a local minimum during the experimental run, unable to converge to the correct local minimum. 

A key observation is the presence of repeated patterns in the stored set. Specifically, the last two stored patterns are repetitions of the first two classes but with different instances (e.g., class $0$ appears twice, each with distinct shapes). These repeated patterns exhibit high cross-correlation values due to their structural similarity. For instance, the cross-correlation between $\xi^{\scriptscriptstyle(1)}$ (the first instance of digit $0$) and $\xi^{\scriptscriptstyle(11)}$ (the second instance of digit $0$) is $\rho_{1,11} = \rho_{11,1} = 0.70$, significantly higher than the average correlation $\bar{\rho} \approx 0.22$. Despite this high correlation, both patterns $\xi^{\scriptscriptstyle(1)}$ and $\xi^{\scriptscriptstyle(11)}$ are successfully retrieved, demonstrating remarkable robustness against sharp noise fluctuations.

\subsection{Retrieval beyond Critical Capacity ($K > K_c$)}\label{Retrieval beyond Critical Capacity}

\begin{figure}[H]
    \centering
    \includegraphics[width=\linewidth]{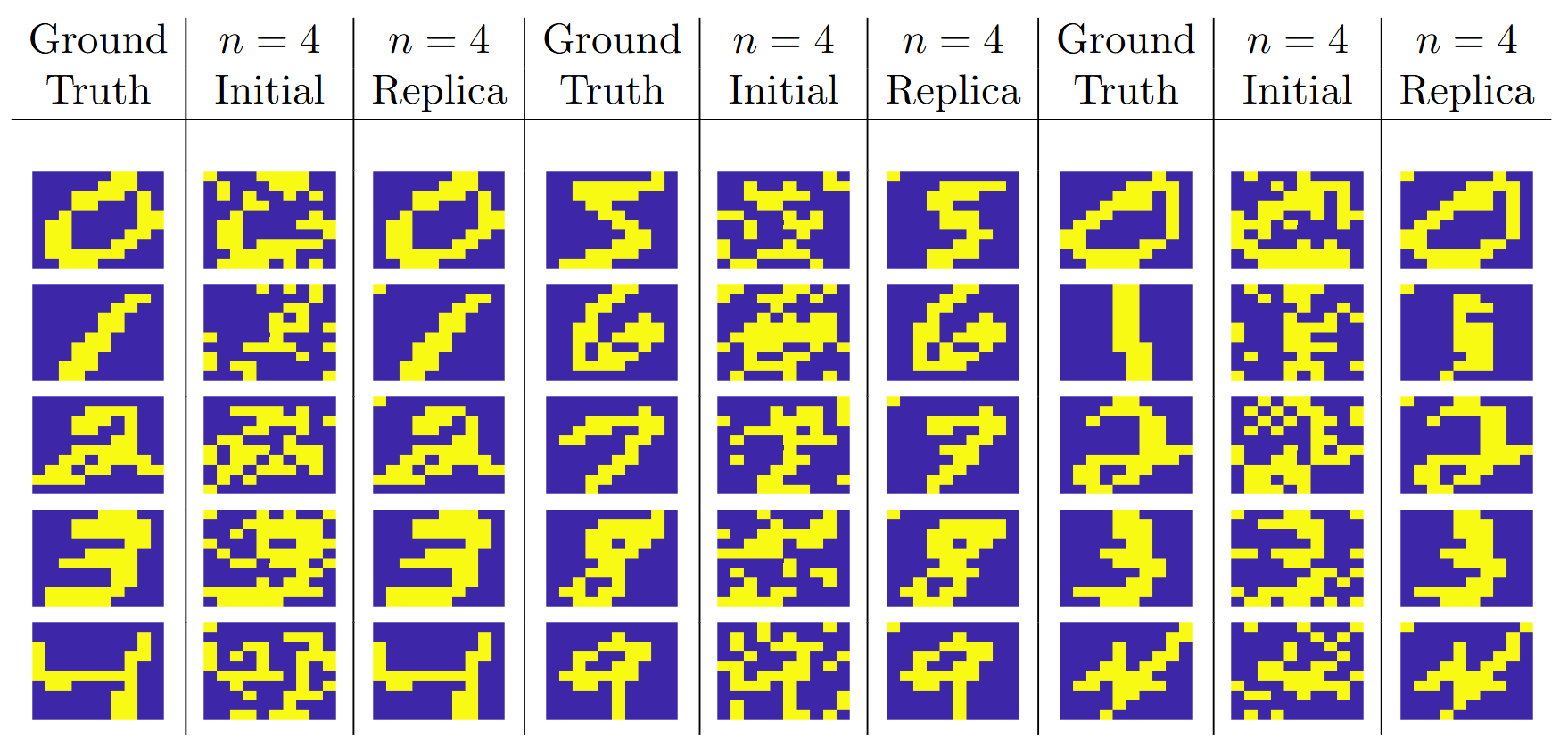}
    
\caption{{\bf Retrieval at Critical Capacity ($K > K_c$).} MNIST digit pattern retrieval performance for pattern capacity $K = 15$ using the 4-body interaction. Patterns $\xi^{\scriptscriptstyle(6)}$ and $\xi^{\scriptscriptstyle(12)}$ have converged to a mixed state of patterns, which yields to a retrieval accuracy value of $\eta \approx 0.87$.}
    \label{fig:stored_patterns_3}
\end{figure}

Above critical capacity $K > K_c$, we observe a continuous decrease in retrieval accuracy. For $K = 15$, the retrieval accuracy decreases to $\eta \approx 0.87$. This behavior is shown in Fig.~\ref{fig:stored_patterns_3}, where more replicas are either mixed or incorrectly recalled due to the system being overwhelmed with the extra stored patterns beyond the limit it can process. 

\newpage

\section{Digitally Emulated Higher-Order Interactions in OHNNs}
\label{app:digital_emulation}
\begin{figure}[H]
   \begin{center}
       \includegraphics[width=\linewidth]{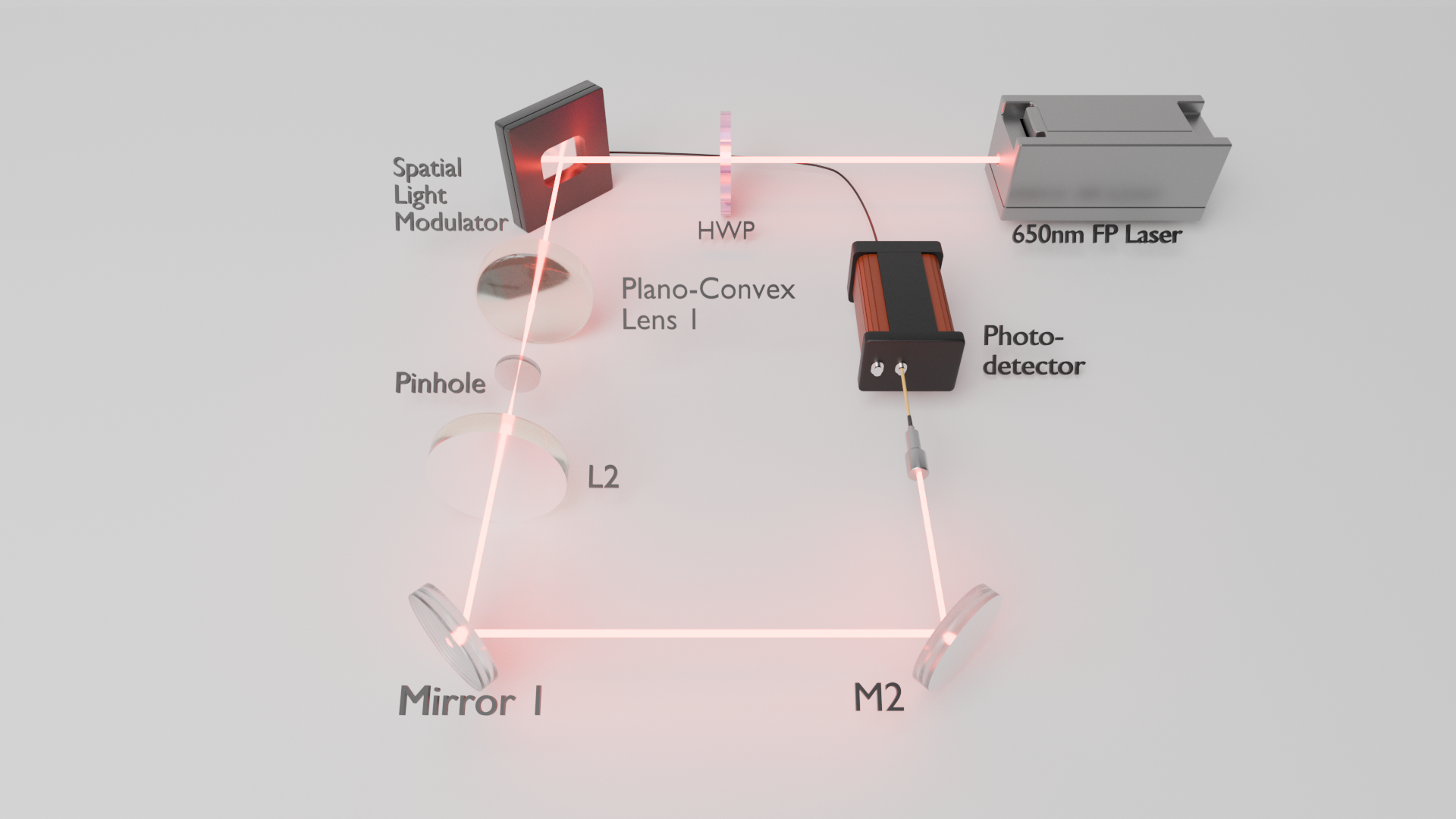} 
  
    \caption{\textbf{Experimental setup for implementing OHNN with digital high-order interactions.} Light is incident on the SLM, which spatially modulates the beam into multiple diffraction orders. A pinhole isolates the first diffraction order, and the signal light is coupled into a single-mode fiber for measurement. The measured intensity \( I \) is digitally processed by raising it to the power \( n/2 \), emulating \( n\text{th} \)-order interactions.}
    \label{fig_exp_hybrid}

   \end{center}
\end{figure}

To further validate our NOHNN findings, we also tested our previous experimental setup, shown in Fig.~\ref{fig_exp_hybrid} and discussed in Ref. \cite{katidis2024robust}, by digitally introducing nonlinearity into the system. Here, we emulate the $n=4$ interaction in OHNNs by digitally squaring optically measured $n=2$ intensities, avoiding physical nonlinearities. While the optical framework (e.g., Fourier-based matrix operations) is retained and critical capacity ($K_c$) and noise robustness improve, the method relies on digital post-processing, introducing latency compared to all-optical systems. Results mirror those of nonlinear-optical implementations (Section~\ref{Results}), validating the scalability of nonlinear-optical higher-order couplings. For latency-free systems, nonlinear components (e.g., SH crystals) remain necessary to eliminate digital overhead. This approach is described by the following energy function:

\begin{equation}
    H = - \sum_{\mu = 1}^{K} F(I_{\text{signal}}),
    \label{eqD1}
\end{equation}

\noindent where \(F(x) = x^{(n/2)}\) and \(n\) denotes the order of interaction. Theoretically, this system can model any even-order interaction ($n = 2,4,6,\dots$), but for comparison purposes, we limit our analysis to $n = 2$ and $4$. The rest of this section will show results using this hybrid system for the same types of patterns shown in Section~\ref{Results}.

\newpage

\subsection{Hadamard Patterns}

\begin{figure}[H]
   \centering
       \includegraphics[width=0.9\linewidth]{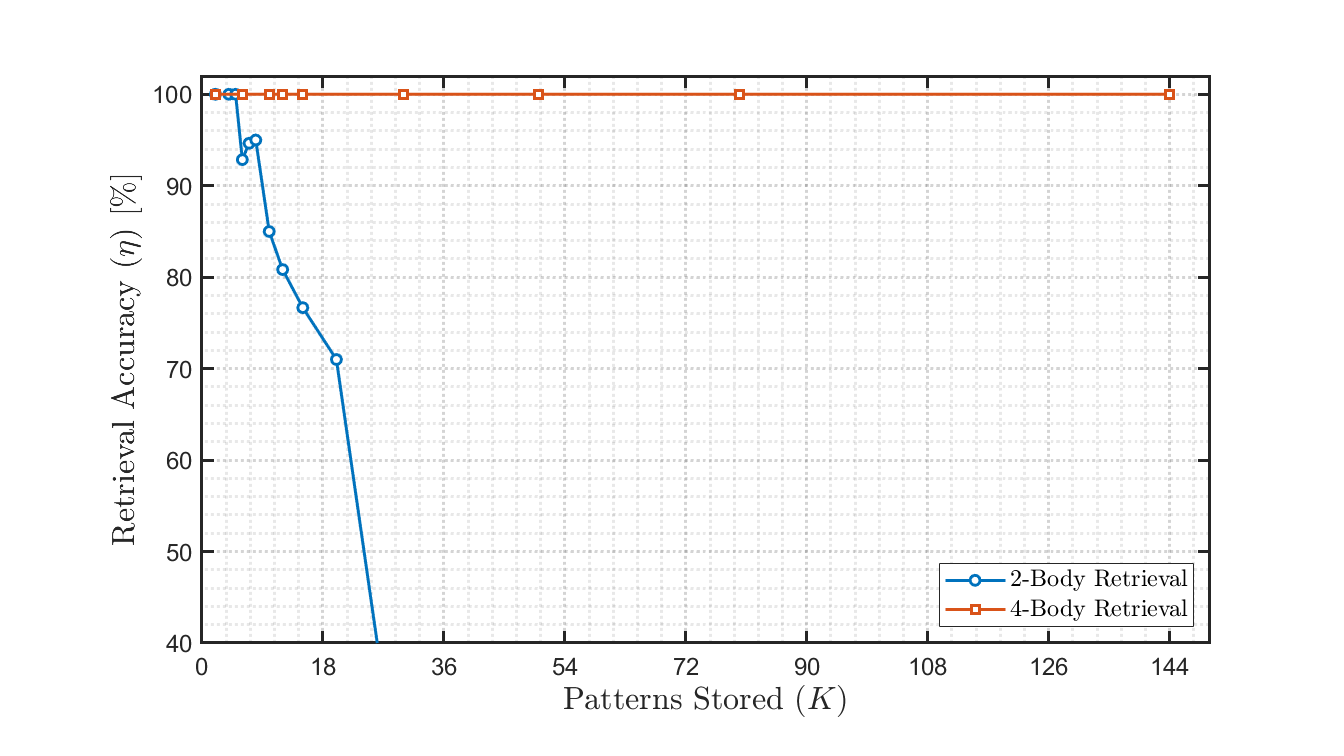} 

    \caption{{\bf Retrieval accuracy of 4-Body vs. 2-Body on Hadamard Patterns.} Retrieval accuracy of traditional AM (blue) and DAM (red) as a function of $K$, using Hadamard patterns. The plot demonstrates a minimum pattern capacity improvement of ${\sim}28.8$ times, with $K_c^{\scriptscriptstyle (n = 4)} \geq 144$ for the 4-body interaction model, compared to $K_c^{\scriptscriptstyle (n = 2)} = 5$ for the traditional 2-body model.}
    \label{fig_had_hybrid}

\end{figure}

To evaluate retrieval accuracy $\eta$, we set the similarity threshold $\tau = 0.45$ to maintain sufficient separation from orthogonal patterns. For these runs, the number of neurons used is N = $72$. As shown in Fig.~\ref{fig_had_hybrid}, increasing the interaction order boosts the critical pattern capacity \(K_{c}\) by at least a factor of~28 (\(K_c^{\scriptscriptstyle (n=4)} / K_c^{\scriptscriptstyle (n=2)} \geq 28.8\)). While the 4-body critical pattern capacity upper bound remains undetermined, tests with larger $K$ values could be performed to identify this limit. \bigskip

\begin{figure}[H]
   \centering
       \includegraphics[width=1\linewidth]{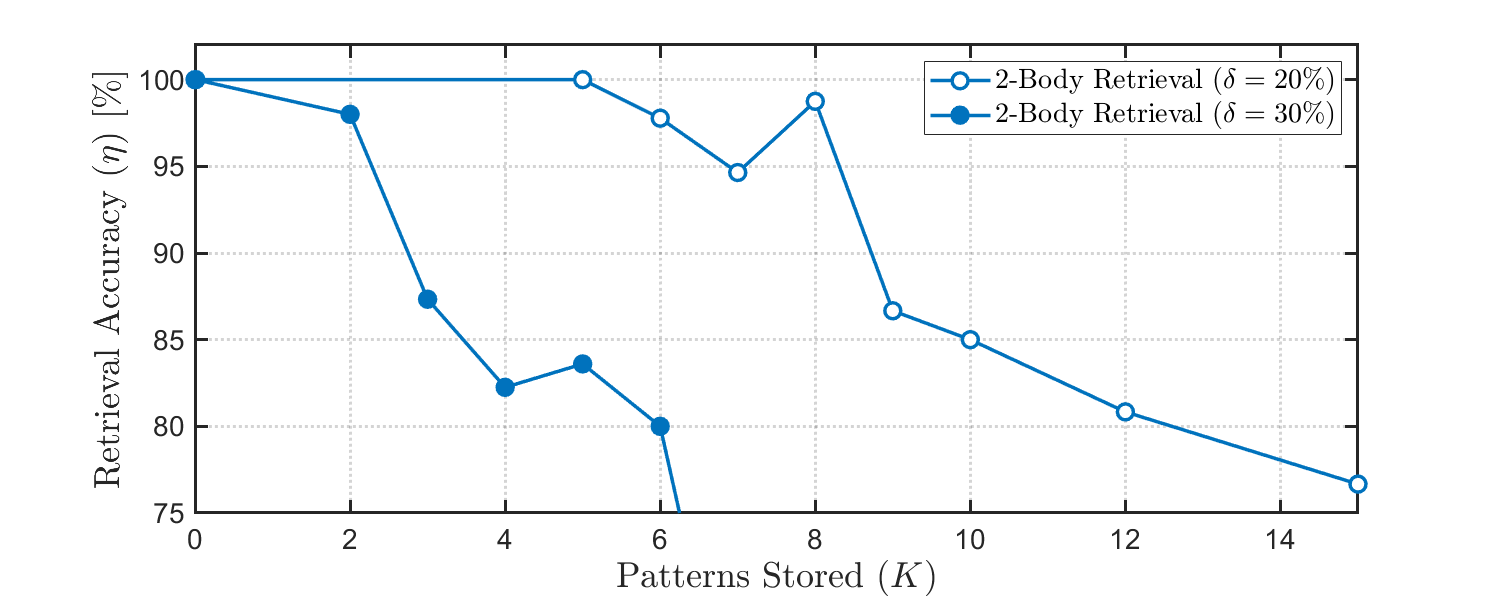} \\
       \includegraphics[width=1\linewidth]{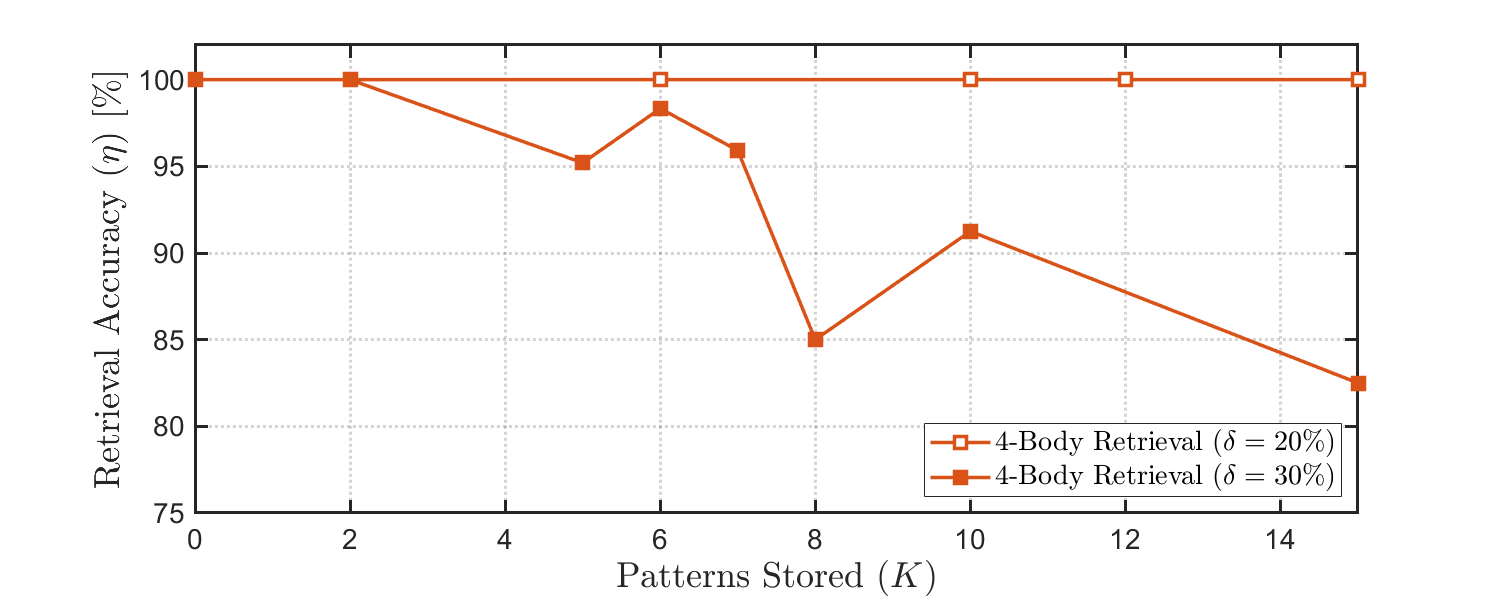} 

    \caption{{\bf Retrieval accuracy performance varying $\delta$ on Hadamard Patterns.} Retrieval accuracy of traditional associative memory (AM, blue) and modern DAM (red) as a function of~\( K \), for varying initial masking percentages~\( \delta \). The results illustrate how masking impacts pattern storage capacity across energy models. At~\( \delta = 0.30 \), storage capacity remains similar for both models, whereas at~\( \delta = 0.20 \), a clear divergence emerges, with DAM supporting higher capacities. }
    \label{fig_had_delta_hybrid}

\end{figure}

 Fig.~\ref{fig_had_delta_hybrid} further analyzes how masking percentage \(\delta\) affects retrieval across energy models. As $\delta$ decreases, both 2-body and 4-body systems are able to store a higher number of patterns. At $\delta = 0.20$, a clear distinction in $K_c$ can be made between the two energy model. For example, the 4-body model shows that $K_c^{(n=4)}>14$, while the 2-body model gives $K_c^{(n=2)}=5$. This distinction diminishes when the masking is increased to $\delta=0.30$. In this case, the critical pattern capacities of both models are very similar, with the 4-body model's retrieval performance being slightly better compared to that of the 2-body. 

\newpage             

\subsection{Correlated Random Patterns}

\begin{figure}[H]
    \centering
   \includegraphics[width=0.75\linewidth]{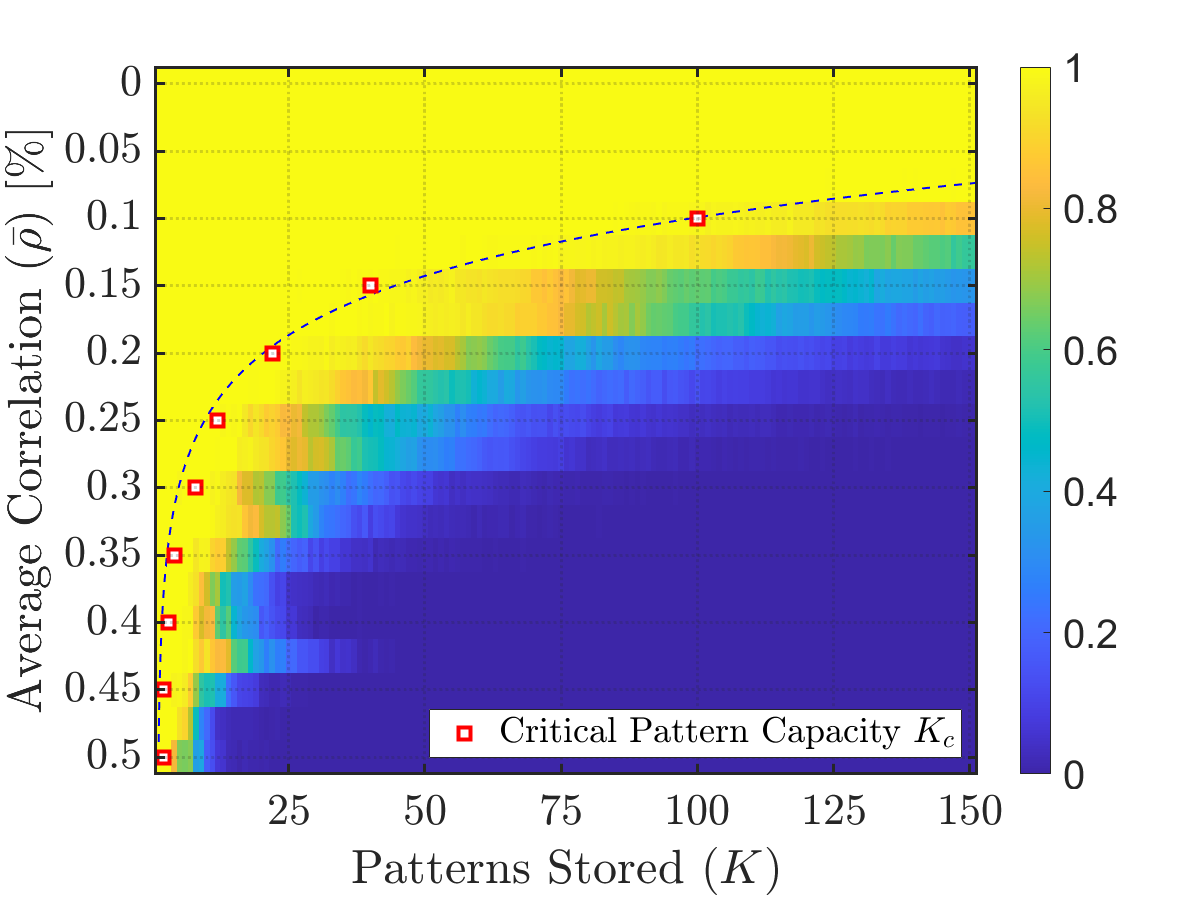} 
   \label{fig:4_body_phase_transition_hybrid}
   
   \caption{{\bf Pattern storage capacity as a function of average correlation for 2-body and 4-body interactions.} Simulation color maps for the number of patterns stored \(K\) as a function of average correlation \(\bar{\rho}\), demonstrating a distinct phase transition curve. The energy model uses the hybrid model in Eq.~(\ref{eqD1}) with $n = 4$. The color bar represents the retrieval accuracy of the system, averaged over 50 trials per \((K, \bar{\rho})\) parameter set. Experimental data points (red open squares), representing the critical capacity \(K_{c}\) for each correlation value, are overlaid to validate the theoretical predictions and study the matching behavior.}
   \label{fig_phase_transition_hybrid}
\end{figure}

Fig.~\ref{fig_phase_transition_hybrid} shows a color map simulation of the retrieval accuracy with different number of stored patterns and average correlations $(K,\bar{\rho})$. Real experimental data in red are collected from the hybrid system and plotted over the color map. The data describe the critical pattern capacity $K_c$ for different average correlation values. To quantify the results, the experimental data for the 4-body model is fitted to the same function in Eq.~(\ref{fitting}). The fitting yields $C_1 = 494.8 \pm 112.7$ and $C_2 = 16.20 \pm 1.970$ with 95\% confidence. This suggests an analytical bound for the hybrid 4-body model's critical pattern capacity: $K_{c}^{(n=4)} = 494$ (at $\bar{\rho} = 0$). This shows improvements up to 49 times compared the 2-body model. These results show similar behavior in retrieval and storage capacity as the NOHNN using nonlinear-optical effects rather than digital emulation to arise the 4-body interaction term in the energy model. \bigskip

\subsection{MNIST Handwritten Digit Patterns}

\begin{figure}[H]
   \centering
       \includegraphics[width=\linewidth]{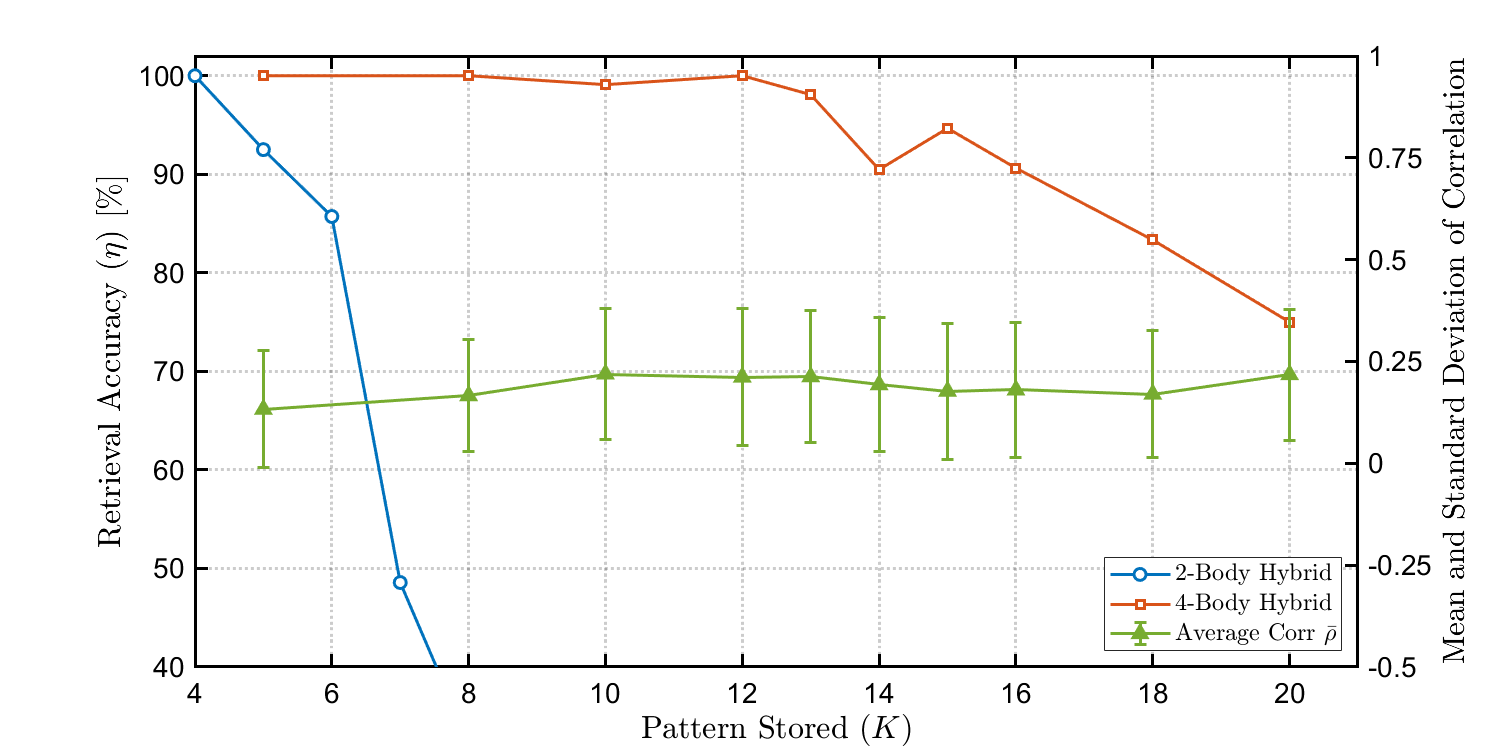} 

    \caption{{\bf Retrieval accuracy of 4-Body vs. 2-Body on MNIST digit Patterns.} Retrieval accuracy $\eta$ (left y-axis) of traditional AM (blue) and hybrid DAM (red) as a function of $K$, using MNIST digit patterns. The plot demonstrates a critical pattern capacity ratio of \( K_c^{\scriptscriptstyle (n = 4)} / K_c^{\scriptscriptstyle (n = 2)} = 3\), allowing for a three times increase in $K_c$. The average correlation \( \bar{\rho} \) and standard deviation (right y-axis) are also shown, highlighting the system's robustness to random noise and fluctuations.}
    \label{fig_MNIST_hybrid}

\end{figure}

Fig.~\ref{fig_MNIST_hybrid} shows \(\eta\) as a function of the number of \(K\) for both the 2-body and 4-body energy models. The results reveal a critical capacity ratio \(K_c^{\scriptscriptstyle (n = 4)} / K_c^{\scriptscriptstyle (n = 2)} = 3\), demonstrating again the superior scalability of the 4-body model. Additionally, the large average correlation \(\bar{\rho}\) and its standard deviation (right y-axis) in those patterns showcase the system's robustness to noise fluctuations caused by pattern correlations. The hybrid NOHNN results show similar behavior in retrieval and storage capacity as the model using nonlinear-optical effects, although the pattern capacity ratio is lower than that of the optical NOHNN. \bigskip

\subsection{NOHNN and hybrid OHNN Comparison}

The results of the digitally emulated nonlinearity model are comparable to those of the NOHNN, yielding very similar performance and thus validating the use of nonlinear optics to achieve the 4-body interaction. Starting with Hadamard patterns, the digitally emulated system outperformed the NOHNN, primarily due to testing with a higher number of patterns (\(K_{\text{max}}^{(\text{hybrid})} = 144\) vs.\ \(K_{\text{max}}^{(\text{NOHNN})} = 50\)). Consequently, we conclude that the 4-body interaction enhances the critical pattern capacity \(K_c\) by a factor of at least \(28.8\). Neither models exhibit retrieval failure even at \(K_{\text{max}}\), suggesting that the critical capacity ratio could improve further with tests using larger \(K\). This trend persists with correlated and MNIST patterns, though the NOHNN demonstrates marginally superior performance in these cases. This highlights the robustness and superiority of the optical nonlinear effects role in physical implementations, which could capture complex higher-order interactions better compared to digital emulation.




\end{appendices}


\bibliography{Main}% common bib file


\begin{thebibliography}{72}
\ifx \bisbn   \undefined \def \bisbn  #1{ISBN #1}\fi
\ifx \binits  \undefined \def \binits#1{#1}\fi
\ifx \bauthor  \undefined \def \bauthor#1{#1}\fi
\ifx \batitle  \undefined \def \batitle#1{#1}\fi
\ifx \bjtitle  \undefined \def \bjtitle#1{#1}\fi
\ifx \bvolume  \undefined \def \bvolume#1{\textbf{#1}}\fi
\ifx \byear  \undefined \def \byear#1{#1}\fi
\ifx \bissue  \undefined \def \bissue#1{#1}\fi
\ifx \bfpage  \undefined \def \bfpage#1{#1}\fi
\ifx \blpage  \undefined \def \blpage #1{#1}\fi
\ifx \burl  \undefined \def \burl#1{\textsf{#1}}\fi
\ifx \doiurl  \undefined \def \doiurl#1{\url{https://doi.org/#1}}\fi
\ifx \betal  \undefined \def \betal{\textit{et al.}}\fi
\ifx \binstitute  \undefined \def \binstitute#1{#1}\fi
\ifx \binstitutionaled  \undefined \def \binstitutionaled#1{#1}\fi
\ifx \bctitle  \undefined \def \bctitle#1{#1}\fi
\ifx \beditor  \undefined \def \beditor#1{#1}\fi
\ifx \bpublisher  \undefined \def \bpublisher#1{#1}\fi
\ifx \bbtitle  \undefined \def \bbtitle#1{#1}\fi
\ifx \bedition  \undefined \def \bedition#1{#1}\fi
\ifx \bseriesno  \undefined \def \bseriesno#1{#1}\fi
\ifx \blocation  \undefined \def \blocation#1{#1}\fi
\ifx \bsertitle  \undefined \def \bsertitle#1{#1}\fi
\ifx \bsnm \undefined \def \bsnm#1{#1}\fi
\ifx \bsuffix \undefined \def \bsuffix#1{#1}\fi
\ifx \bparticle \undefined \def \bparticle#1{#1}\fi
\ifx \barticle \undefined \def \barticle#1{#1}\fi
\bibcommenthead
\ifx \bconfdate \undefined \def \bconfdate #1{#1}\fi
\ifx \botherref \undefined \def \botherref #1{#1}\fi
\ifx \url \undefined \def \url#1{\textsf{#1}}\fi
\ifx \bchapter \undefined \def \bchapter#1{#1}\fi
\ifx \bbook \undefined \def \bbook#1{#1}\fi
\ifx \bcomment \undefined \def \bcomment#1{#1}\fi
\ifx \oauthor \undefined \def \oauthor#1{#1}\fi
\ifx \citeauthoryear \undefined \def \citeauthoryear#1{#1}\fi
\ifx \endbibitem  \undefined \def \endbibitem {}\fi
\ifx \bconflocation  \undefined \def \bconflocation#1{#1}\fi
\ifx \arxivurl  \undefined \def \arxivurl#1{\textsf{#1}}\fi
\csname PreBibitemsHook\endcsname

%%% 1
\bibitem[\protect\citeauthoryear{Katz and Frost}{1996}]{katz1996intrinsic}
\begin{barticle}
\bauthor{\bsnm{Katz}, \binits{P.S.}},
\bauthor{\bsnm{Frost}, \binits{W.N.}}:
\batitle{Intrinsic neuromodulation: altering neuronal circuits from within}.
\bjtitle{Trends in neurosciences}
\bvolume{19}(\bissue{2}),
\bfpage{54}--\blpage{61}
(\byear{1996})
\end{barticle}
\endbibitem

%%% 2
\bibitem[\protect\citeauthoryear{Katz et~al.}{1994}]{katz1994dynamic}
\begin{barticle}
\bauthor{\bsnm{Katz}, \binits{P.S.}},
\bauthor{\bsnm{Getting}, \binits{P.A.}},
\bauthor{\bsnm{Frost}, \binits{W.N.}}:
\batitle{Dynamic neuromodulation of synaptic strength intrinsic to a central pattern generator circuit}.
\bjtitle{Nature}
\bvolume{367}(\bissue{6465}),
\bfpage{729}--\blpage{731}
(\byear{1994})
\end{barticle}
\endbibitem

%%% 3
\bibitem[\protect\citeauthoryear{Schurr et~al.}{2024}]{Schurr2024}
\begin{barticle}
\bauthor{\bsnm{Schurr}, \binits{R.}},
\bauthor{\bsnm{Reznik}, \binits{D.}},
\bauthor{\bsnm{Hillman}, \binits{H.}},
\bauthor{\bsnm{Bhui}, \binits{R.}},
\bauthor{\bsnm{Gershman}, \binits{S.J.}}:
\batitle{Dynamic computational phenotyping of human cognition}.
\bjtitle{Nature Human Behaviour}
\bvolume{8}(\bissue{5}),
\bfpage{917}--\blpage{931}
(\byear{2024})
\doiurl{10.1038/s41562-024-01814-x}
\end{barticle}
\endbibitem

%%% 4
\bibitem[\protect\citeauthoryear{Hopfield}{1982}]{hopfield1982neural}
\begin{barticle}
\bauthor{\bsnm{Hopfield}, \binits{J.J.}}:
\batitle{Neural networks and physical systems with emergent collective computational abilities.}
\bjtitle{Proceedings of the national academy of sciences}
\bvolume{79}(\bissue{8}),
\bfpage{2554}--\blpage{2558}
(\byear{1982})
\end{barticle}
\endbibitem

%%% 5
\bibitem[\protect\citeauthoryear{Suganthan et~al.}{1995}]{suganthan1995pattern}
\begin{barticle}
\bauthor{\bsnm{Suganthan}, \binits{P.N.}},
\bauthor{\bsnm{Teoh}, \binits{E.K.}},
\bauthor{\bsnm{Mital}, \binits{D.P.}}:
\batitle{Pattern recognition by homomorphic graph matching using hopfield neural networks}.
\bjtitle{Image and Vision Computing}
\bvolume{13}(\bissue{1}),
\bfpage{45}--\blpage{60}
(\byear{1995})
\end{barticle}
\endbibitem

%%% 6
\bibitem[\protect\citeauthoryear{Nasrabadi and Li}{1991}]{nasrabadi1991object}
\begin{barticle}
\bauthor{\bsnm{Nasrabadi}, \binits{N.M.}},
\bauthor{\bsnm{Li}, \binits{W.}}:
\batitle{Object recognition by a hopfield neural network}.
\bjtitle{IEEE Transactions on Systems, Man, and Cybernetics}
\bvolume{21}(\bissue{6}),
\bfpage{1523}--\blpage{1535}
(\byear{1991})
\end{barticle}
\endbibitem

%%% 7
\bibitem[\protect\citeauthoryear{Tank and Hopfield}{1986}]{tank1986simple}
\begin{barticle}
\bauthor{\bsnm{Tank}, \binits{D.}},
\bauthor{\bsnm{Hopfield}, \binits{J.}}:
\batitle{Simple'neural'optimization networks: An a/d converter, signal decision circuit, and a linear programming circuit}.
\bjtitle{IEEE transactions on circuits and systems}
\bvolume{33}(\bissue{5}),
\bfpage{533}--\blpage{541}
(\byear{1986})
\end{barticle}
\endbibitem

%%% 8
\bibitem[\protect\citeauthoryear{Hopfield and Tank}{1985}]{hopfield1985neural}
\begin{barticle}
\bauthor{\bsnm{Hopfield}, \binits{J.J.}},
\bauthor{\bsnm{Tank}, \binits{D.W.}}:
\batitle{“neural” computation of decisions in optimization problems}.
\bjtitle{Biological cybernetics}
\bvolume{52}(\bissue{3}),
\bfpage{141}--\blpage{152}
(\byear{1985})
\end{barticle}
\endbibitem

%%% 9
\bibitem[\protect\citeauthoryear{He et~al.}{2022}]{he2022masked}
\begin{bchapter}
\bauthor{\bsnm{He}, \binits{K.}},
\bauthor{\bsnm{Chen}, \binits{X.}},
\bauthor{\bsnm{Xie}, \binits{S.}},
\bauthor{\bsnm{Li}, \binits{Y.}},
\bauthor{\bsnm{Doll{\'a}r}, \binits{P.}},
\bauthor{\bsnm{Girshick}, \binits{R.}}:
\bctitle{Masked autoencoders are scalable vision learners}.
In: \bbtitle{Proceedings of the IEEE/CVF Conference on Computer Vision and Pattern Recognition},
pp. \bfpage{16000}--\blpage{16009}
(\byear{2022})
\end{bchapter}
\endbibitem

%%% 10
\bibitem[\protect\citeauthoryear{Agliari and De~Marzo}{2020}]{agliari2020tolerance}
\begin{barticle}
\bauthor{\bsnm{Agliari}, \binits{E.}},
\bauthor{\bsnm{De~Marzo}, \binits{G.}}:
\batitle{Tolerance versus synaptic noise in dense associative memories}.
\bjtitle{The European Physical Journal Plus}
\bvolume{135}(\bissue{11}),
\bfpage{1}--\blpage{22}
(\byear{2020})
\end{barticle}
\endbibitem

%%% 11
\bibitem[\protect\citeauthoryear{Alonso and Krichmar}{2024}]{alonso2024sparse}
\begin{barticle}
\bauthor{\bsnm{Alonso}, \binits{N.}},
\bauthor{\bsnm{Krichmar}, \binits{J.L.}}:
\batitle{A sparse quantized hopfield network for online-continual memory}.
\bjtitle{Nature Communications}
\bvolume{15}(\bissue{1}),
\bfpage{3722}
(\byear{2024})
\end{barticle}
\endbibitem

%%% 12
\bibitem[\protect\citeauthoryear{Hu et~al.}{2015}]{hu2015associative}
\begin{barticle}
\bauthor{\bsnm{Hu}, \binits{S.}},
\bauthor{\bsnm{Liu}, \binits{Y.}},
\bauthor{\bsnm{Liu}, \binits{Z.}},
\bauthor{\bsnm{Chen}, \binits{T.}},
\bauthor{\bsnm{Wang}, \binits{J.}},
\bauthor{\bsnm{Yu}, \binits{Q.}},
\bauthor{\bsnm{Deng}, \binits{L.}},
\bauthor{\bsnm{Yin}, \binits{Y.}},
\bauthor{\bsnm{Hosaka}, \binits{S.}}:
\batitle{Associative memory realized by a reconfigurable memristive hopfield neural network}.
\bjtitle{Nature communications}
\bvolume{6}(\bissue{1}),
\bfpage{7522}
(\byear{2015})
\end{barticle}
\endbibitem

%%% 13
\bibitem[\protect\citeauthoryear{Sourlas}{1989}]{sourlas1989spin}
\begin{barticle}
\bauthor{\bsnm{Sourlas}, \binits{N.}}:
\batitle{Spin-glass models as error-correcting codes}.
\bjtitle{Nature}
\bvolume{339}(\bissue{6227}),
\bfpage{693}--\blpage{695}
(\byear{1989})
\end{barticle}
\endbibitem

%%% 14
\bibitem[\protect\citeauthoryear{Paik and Katsaggelos}{1992}]{paik1992image}
\begin{barticle}
\bauthor{\bsnm{Paik}, \binits{J.K.}},
\bauthor{\bsnm{Katsaggelos}, \binits{A.K.}}:
\batitle{Image restoration using a modified hopfield network}.
\bjtitle{IEEE Transactions on image processing}
\bvolume{1}(\bissue{1}),
\bfpage{49}--\blpage{63}
(\byear{1992})
\end{barticle}
\endbibitem

%%% 15
\bibitem[\protect\citeauthoryear{Horowitz}{2014}]{6757323}
\begin{bchapter}
\bauthor{\bsnm{Horowitz}, \binits{M.}}:
\bctitle{1.1 computing's energy problem (and what we can do about it)}.
In: \bbtitle{2014 IEEE International Solid-State Circuits Conference Digest of Technical Papers (ISSCC)},
pp. \bfpage{10}--\blpage{14}
(\byear{2014}).
\doiurl{10.1109/ISSCC.2014.6757323}
\end{bchapter}
\endbibitem

%%% 16
\bibitem[\protect\citeauthoryear{Wetzstein et~al.}{2020}]{Wetzstein2020}
\begin{barticle}
\bauthor{\bsnm{Wetzstein}, \binits{G.}},
\bauthor{\bsnm{Ozcan}, \binits{A.}},
\bauthor{\bsnm{Gigan}, \binits{S.}},
\bauthor{\bsnm{Fan}, \binits{S.}},
\bauthor{\bsnm{Englund}, \binits{D.}},
\bauthor{\bsnm{Solja{\v{c}}i{\'{c}}}, \binits{M.}},
\bauthor{\bsnm{Denz}, \binits{C.}},
\bauthor{\bsnm{Miller}, \binits{D.A.B.}},
\bauthor{\bsnm{Psaltis}, \binits{D.}}:
\batitle{Inference in artificial intelligence with deep optics and photonics}.
\bjtitle{Nature}
\bvolume{588}(\bissue{7836}),
\bfpage{39}--\blpage{47}
(\byear{2020})
\doiurl{10.1038/s41586-020-2973-6}
\end{barticle}
\endbibitem

%%% 17
\bibitem[\protect\citeauthoryear{Shastri et~al.}{2021}]{shastri2021photonics}
\begin{barticle}
\bauthor{\bsnm{Shastri}, \binits{B.J.}},
\bauthor{\bsnm{Tait}, \binits{A.N.}},
\bauthor{\bsnm{Lima}, \binits{T.}},
\bauthor{\bsnm{Pernice}, \binits{W.H.}},
\bauthor{\bsnm{Bhaskaran}, \binits{H.}},
\bauthor{\bsnm{Wright}, \binits{C.D.}},
\bauthor{\bsnm{Prucnal}, \binits{P.R.}}:
\batitle{Photonics for artificial intelligence and neuromorphic computing}.
\bjtitle{Nature Photonics}
\bvolume{15}(\bissue{2}),
\bfpage{102}--\blpage{114}
(\byear{2021})
\end{barticle}
\endbibitem

%%% 18
\bibitem[\protect\citeauthoryear{Wang et~al.}{2024}]{wang2024optical}
\begin{botherref}
\oauthor{\bsnm{Wang}, \binits{Z.}},
\oauthor{\bsnm{M{\"u}ller}, \binits{K.}},
\oauthor{\bsnm{Filipovich}, \binits{M.}},
\oauthor{\bsnm{Launay}, \binits{J.}},
\oauthor{\bsnm{Ohana}, \binits{R.}},
\oauthor{\bsnm{Pariente}, \binits{G.}},
\oauthor{\bsnm{Mokaadi}, \binits{S.}},
\oauthor{\bsnm{Brossollet}, \binits{C.}},
\oauthor{\bsnm{Moreau}, \binits{F.}},
\oauthor{\bsnm{Cappelli}, \binits{A.}}, et al.:
Optical training of large-scale transformers and deep neural networks with direct feedback alignment.
arXiv preprint arXiv:2409.12965
(2024)
\end{botherref}
\endbibitem

%%% 19
\bibitem[\protect\citeauthoryear{Anderson et~al.}{2023}]{anderson2023optical}
\begin{botherref}
\oauthor{\bsnm{Anderson}, \binits{M.}},
\oauthor{\bsnm{Ma}, \binits{S.-Y.}},
\oauthor{\bsnm{Wang}, \binits{T.}},
\oauthor{\bsnm{Wright}, \binits{L.}},
\oauthor{\bsnm{McMahon}, \binits{P.}}:
Optical transformers.
Transactions on Machine Learning Research
(2023)
\end{botherref}
\endbibitem

%%% 20
\bibitem[\protect\citeauthoryear{Leonetti et~al.}{2024}]{leonetti2024photonic}
\begin{barticle}
\bauthor{\bsnm{Leonetti}, \binits{M.}},
\bauthor{\bsnm{Gosti}, \binits{G.}},
\bauthor{\bsnm{Ruocco}, \binits{G.}}:
\batitle{Photonic stochastic emergent storage for deep classification by scattering-intrinsic patterns}.
\bjtitle{Nature Communications}
\bvolume{15}(\bissue{1}),
\bfpage{505}
(\byear{2024})
\end{barticle}
\endbibitem

%%% 21
\bibitem[\protect\citeauthoryear{Kumar et~al.}{2021}]{kumar2021robust}
\begin{barticle}
\bauthor{\bsnm{Kumar}, \binits{S.}},
\bauthor{\bsnm{Bu}, \binits{T.}},
\bauthor{\bsnm{Zhang}, \binits{H.}},
\bauthor{\bsnm{Huang}, \binits{I.}},
\bauthor{\bsnm{Huang}, \binits{Y.}}:
\batitle{Robust and efficient single-pixel image classification with nonlinear optics}.
\bjtitle{Optics Letters}
\bvolume{46}(\bissue{8}),
\bfpage{1848}--\blpage{1851}
(\byear{2021})
\end{barticle}
\endbibitem

%%% 22
\bibitem[\protect\citeauthoryear{Xu et~al.}{2024}]{xu2024large}
\begin{barticle}
\bauthor{\bsnm{Xu}, \binits{Z.}},
\bauthor{\bsnm{Zhou}, \binits{T.}},
\bauthor{\bsnm{Ma}, \binits{M.}},
\bauthor{\bsnm{Deng}, \binits{C.}},
\bauthor{\bsnm{Dai}, \binits{Q.}},
\bauthor{\bsnm{Fang}, \binits{L.}}:
\batitle{Large-scale photonic chiplet taichi empowers 160-tops/w artificial general intelligence}.
\bjtitle{Science}
\bvolume{384}(\bissue{6692}),
\bfpage{202}--\blpage{209}
(\byear{2024})
\end{barticle}
\endbibitem

%%% 23
\bibitem[\protect\citeauthoryear{Ramsauer et~al.}{2020}]{ramsauer2020hopfield}
\begin{botherref}
\oauthor{\bsnm{Ramsauer}, \binits{H.}},
\oauthor{\bsnm{Sch{\"a}fl}, \binits{B.}},
\oauthor{\bsnm{Lehner}, \binits{J.}},
\oauthor{\bsnm{Seidl}, \binits{P.}},
\oauthor{\bsnm{Widrich}, \binits{M.}},
\oauthor{\bsnm{Adler}, \binits{T.}},
\oauthor{\bsnm{Gruber}, \binits{L.}},
\oauthor{\bsnm{Holzleitner}, \binits{M.}},
\oauthor{\bsnm{Pavlovi{\'c}}, \binits{M.}},
\oauthor{\bsnm{Sandve}, \binits{G.K.}}, et al.:
Hopfield networks is all you need.
arXiv preprint arXiv:2008.02217
(2020)
\end{botherref}
\endbibitem

%%% 24
\bibitem[\protect\citeauthoryear{Krotov and Hopfield}{2020}]{krotov2020large}
\begin{botherref}
\oauthor{\bsnm{Krotov}, \binits{D.}},
\oauthor{\bsnm{Hopfield}, \binits{J.}}:
Large associative memory problem in neurobiology and machine learning.
arXiv preprint arXiv:2008.06996
(2020)
\end{botherref}
\endbibitem

%%% 25
\bibitem[\protect\citeauthoryear{Denz}{2013}]{denz2013optical}
\begin{bbook}
\bauthor{\bsnm{Denz}, \binits{C.}}:
\bbtitle{Optical Neural Networks}.
\bsertitle{Optics and photonics}.
\bpublisher{Vieweg+Teubner Verlag}, \blocation{???}
(\byear{2013}).
\burl{https://books.google.com/books?id=QMiqCAAAQBAJ}
\end{bbook}
\endbibitem

%%% 26
\bibitem[\protect\citeauthoryear{Bausch and Leditzky}{2020}]{bausch2020quantum}
\begin{barticle}
\bauthor{\bsnm{Bausch}, \binits{J.}},
\bauthor{\bsnm{Leditzky}, \binits{F.}}:
\batitle{Quantum codes from neural networks}.
\bjtitle{New Journal of Physics}
\bvolume{22}(\bissue{2}),
\bfpage{023005}
(\byear{2020})
\end{barticle}
\endbibitem

%%% 27
\bibitem[\protect\citeauthoryear{Marsh et~al.}{2021}]{marsh2021enhancing}
\begin{barticle}
\bauthor{\bsnm{Marsh}, \binits{B.P.}},
\bauthor{\bsnm{Guo}, \binits{Y.}},
\bauthor{\bsnm{Kroeze}, \binits{R.M.}},
\bauthor{\bsnm{Gopalakrishnan}, \binits{S.}},
\bauthor{\bsnm{Ganguli}, \binits{S.}},
\bauthor{\bsnm{Keeling}, \binits{J.}},
\bauthor{\bsnm{Lev}, \binits{B.L.}}:
\batitle{Enhancing associative memory recall and storage capacity using confocal cavity qed}.
\bjtitle{Physical Review X}
\bvolume{11}(\bissue{2}),
\bfpage{021048}
(\byear{2021})
\end{barticle}
\endbibitem

%%% 28
\bibitem[\protect\citeauthoryear{Negri et~al.}{2023}]{negri2023storage}
\begin{barticle}
\bauthor{\bsnm{Negri}, \binits{M.}},
\bauthor{\bsnm{Lauditi}, \binits{C.}},
\bauthor{\bsnm{Perugini}, \binits{G.}},
\bauthor{\bsnm{Lucibello}, \binits{C.}},
\bauthor{\bsnm{Malatesta}, \binits{E.}}:
\batitle{Storage and learning phase transitions in the random-features hopfield model}.
\bjtitle{Physical Review Letters}
\bvolume{131}(\bissue{25}),
\bfpage{257301}
(\byear{2023})
\end{barticle}
\endbibitem

%%% 29
\bibitem[\protect\citeauthoryear{Amit et~al.}{1985}]{amit1985storing}
\begin{barticle}
\bauthor{\bsnm{Amit}, \binits{D.J.}},
\bauthor{\bsnm{Gutfreund}, \binits{H.}},
\bauthor{\bsnm{Sompolinsky}, \binits{H.}}:
\batitle{Storing infinite numbers of patterns in a spin-glass model of neural networks}.
\bjtitle{Physical Review Letters}
\bvolume{55}(\bissue{14}),
\bfpage{1530}
(\byear{1985})
\end{barticle}
\endbibitem

%%% 30
\bibitem[\protect\citeauthoryear{Demircigil et~al.}{2017}]{demircigil2017model}
\begin{barticle}
\bauthor{\bsnm{Demircigil}, \binits{M.}},
\bauthor{\bsnm{Heusel}, \binits{J.}},
\bauthor{\bsnm{L{\"o}we}, \binits{M.}},
\bauthor{\bsnm{Upgang}, \binits{S.}},
\bauthor{\bsnm{Vermet}, \binits{F.}}:
\batitle{On a model of associative memory with huge storage capacity}.
\bjtitle{Journal of Statistical Physics}
\bvolume{168},
\bfpage{288}--\blpage{299}
(\byear{2017})
\end{barticle}
\endbibitem

%%% 31
\bibitem[\protect\citeauthoryear{Lucibello and M{\'e}zard}{2024}]{lucibello2024exponential}
\begin{barticle}
\bauthor{\bsnm{Lucibello}, \binits{C.}},
\bauthor{\bsnm{M{\'e}zard}, \binits{M.}}:
\batitle{Exponential capacity of dense associative memories}.
\bjtitle{Physical Review Letters}
\bvolume{132}(\bissue{7}),
\bfpage{077301}
(\byear{2024})
\end{barticle}
\endbibitem

%%% 32
\bibitem[\protect\citeauthoryear{Li et~al.}{2024}]{li2024first}
\begin{botherref}
\oauthor{\bsnm{Li}, \binits{L.}},
\oauthor{\bsnm{Kumar}, \binits{S.}},
\oauthor{\bsnm{Garikapati}, \binits{M.}},
\oauthor{\bsnm{Huang}, \binits{Y.-P.}}:
First photon machine learning.
arXiv preprint arXiv:2410.17471
(2024)
\end{botherref}
\endbibitem

%%% 33
\bibitem[\protect\citeauthoryear{Cai et~al.}{2020}]{cai2020power}
\begin{barticle}
\bauthor{\bsnm{Cai}, \binits{F.}},
\bauthor{\bsnm{Kumar}, \binits{S.}},
\bauthor{\bsnm{Van~Vaerenbergh}, \binits{T.}},
\bauthor{\bsnm{Sheng}, \binits{X.}},
\bauthor{\bsnm{Liu}, \binits{R.}},
\bauthor{\bsnm{Li}, \binits{C.}},
\bauthor{\bsnm{Liu}, \binits{Z.}},
\bauthor{\bsnm{Foltin}, \binits{M.}},
\bauthor{\bsnm{Yu}, \binits{S.}},
\bauthor{\bsnm{Xia}, \binits{Q.}}, \betal:
\batitle{Power-efficient combinatorial optimization using intrinsic noise in memristor hopfield neural networks}.
\bjtitle{Nature Electronics}
\bvolume{3}(\bissue{7}),
\bfpage{409}--\blpage{418}
(\byear{2020})
\end{barticle}
\endbibitem

%%% 34
\bibitem[\protect\citeauthoryear{Krotov and Hopfield}{2016}]{krotov2016dense}
\begin{botherref}
\oauthor{\bsnm{Krotov}, \binits{D.}},
\oauthor{\bsnm{Hopfield}, \binits{J.J.}}:
Dense associative memory for pattern recognition.
Advances in neural information processing systems
\textbf{29}
(2016)
\end{botherref}
\endbibitem

%%% 35
\bibitem[\protect\citeauthoryear{Fan et~al.}{2023}]{fan2023photonic}
\begin{barticle}
\bauthor{\bsnm{Fan}, \binits{Z.}},
\bauthor{\bsnm{Lin}, \binits{J.}},
\bauthor{\bsnm{Dai}, \binits{J.}},
\bauthor{\bsnm{Zhang}, \binits{T.}},
\bauthor{\bsnm{Xu}, \binits{K.}}:
\batitle{Photonic hopfield neural network for the ising problem}.
\bjtitle{Optics Express}
\bvolume{31}(\bissue{13}),
\bfpage{21340}--\blpage{21350}
(\byear{2023})
\end{barticle}
\endbibitem

%%% 36
\bibitem[\protect\citeauthoryear{McMahon}{2023}]{mcmahon2023physics}
\begin{barticle}
\bauthor{\bsnm{McMahon}, \binits{P.L.}}:
\batitle{The physics of optical computing}.
\bjtitle{Nature Reviews Physics}
\bvolume{5}(\bissue{12}),
\bfpage{717}--\blpage{734}
(\byear{2023})
\end{barticle}
\endbibitem

%%% 37
\bibitem[\protect\citeauthoryear{Wu et~al.}{2025}]{Wu2025}
\begin{barticle}
\bauthor{\bsnm{Wu}, \binits{T.}},
\bauthor{\bsnm{Li}, \binits{Y.}},
\bauthor{\bsnm{Ge}, \binits{L.}},
\bauthor{\bsnm{Feng}, \binits{L.}}:
\batitle{Field-programmable photonic nonlinearity}.
\bjtitle{Nature Photonics}
(\byear{2025})
\doiurl{10.1038/s41566-025-01660-x}
\end{barticle}
\endbibitem

%%% 38
\bibitem[\protect\citeauthoryear{Katidis et~al.}{2024}]{katidis2024robust}
\begin{barticle}
\bauthor{\bsnm{Katidis}, \binits{M.}},
\bauthor{\bsnm{Musa}, \binits{K.}},
\bauthor{\bsnm{Kumar}, \binits{S.}},
\bauthor{\bsnm{Li}, \binits{Z.}},
\bauthor{\bsnm{Long}, \binits{F.}},
\bauthor{\bsnm{Qu}, \binits{C.}},
\bauthor{\bsnm{Huang}, \binits{Y.-P.}}:
\batitle{Robust pattern retrieval in an optical hopfield neural network}.
\bjtitle{Optics Letters}
\bvolume{50}(\bissue{1}),
\bfpage{225}--\blpage{228}
(\byear{2024})
\end{barticle}
\endbibitem

%%% 39
\bibitem[\protect\citeauthoryear{Spens and Burgess}{2024}]{spens2024generative}
\begin{barticle}
\bauthor{\bsnm{Spens}, \binits{E.}},
\bauthor{\bsnm{Burgess}, \binits{N.}}:
\batitle{A generative model of memory construction and consolidation}.
\bjtitle{Nature human behaviour}
\bvolume{8}(\bissue{3}),
\bfpage{526}--\blpage{543}
(\byear{2024})
\end{barticle}
\endbibitem

%%% 40
\bibitem[\protect\citeauthoryear{Wang et~al.}{2022}]{wang2022optical}
\begin{barticle}
\bauthor{\bsnm{Wang}, \binits{T.}},
\bauthor{\bsnm{Ma}, \binits{S.-Y.}},
\bauthor{\bsnm{Wright}, \binits{L.G.}},
\bauthor{\bsnm{Onodera}, \binits{T.}},
\bauthor{\bsnm{Richard}, \binits{B.C.}},
\bauthor{\bsnm{McMahon}, \binits{P.L.}}:
\batitle{An optical neural network using less than 1 photon per multiplication}.
\bjtitle{Nature Communications}
\bvolume{13}(\bissue{1}),
\bfpage{123}
(\byear{2022})
\end{barticle}
\endbibitem

%%% 41
\bibitem[\protect\citeauthoryear{Shen et~al.}{2017}]{shen2017deep}
\begin{barticle}
\bauthor{\bsnm{Shen}, \binits{Y.}},
\bauthor{\bsnm{Harris}, \binits{N.C.}},
\bauthor{\bsnm{Skirlo}, \binits{S.}},
\bauthor{\bsnm{Prabhu}, \binits{M.}},
\bauthor{\bsnm{Baehr-Jones}, \binits{T.}},
\bauthor{\bsnm{Hochberg}, \binits{M.}},
\bauthor{\bsnm{Sun}, \binits{X.}},
\bauthor{\bsnm{Zhao}, \binits{S.}},
\bauthor{\bsnm{Larochelle}, \binits{H.}},
\bauthor{\bsnm{Englund}, \binits{D.}}, \betal:
\batitle{Deep learning with coherent nanophotonic circuits}.
\bjtitle{Nature photonics}
\bvolume{11}(\bissue{7}),
\bfpage{441}--\blpage{446}
(\byear{2017})
\end{barticle}
\endbibitem

%%% 42
\bibitem[\protect\citeauthoryear{Ma et~al.}{2025}]{ma2025quantum}
\begin{barticle}
\bauthor{\bsnm{Ma}, \binits{S.-Y.}},
\bauthor{\bsnm{Wang}, \binits{T.}},
\bauthor{\bsnm{Laydevant}, \binits{J.}},
\bauthor{\bsnm{Wright}, \binits{L.G.}},
\bauthor{\bsnm{McMahon}, \binits{P.L.}}:
\batitle{Quantum-limited stochastic optical neural networks operating at a few quanta per activation}.
\bjtitle{Nature Communications}
\bvolume{16}(\bissue{1}),
\bfpage{359}
(\byear{2025})
\end{barticle}
\endbibitem

%%% 43
\bibitem[\protect\citeauthoryear{Koh and Takatsuka}{2009}]{koh2009increasing}
\begin{barticle}
\bauthor{\bsnm{Koh}, \binits{Y.W.}},
\bauthor{\bsnm{Takatsuka}, \binits{K.}}:
\batitle{Increasing memory capacity and reducing spurious states in neural networks by introducing coherent and collective firing}.
\bjtitle{Neural computation}
\bvolume{21}(\bissue{5}),
\bfpage{1321}--\blpage{1334}
(\byear{2009})
\end{barticle}
\endbibitem

%%% 44
\bibitem[\protect\citeauthoryear{Bao et~al.}{2022}]{bao2022capacity}
\begin{barticle}
\bauthor{\bsnm{Bao}, \binits{H.}},
\bauthor{\bsnm{Zhang}, \binits{R.}},
\bauthor{\bsnm{Mao}, \binits{Y.}}:
\batitle{The capacity of the dense associative memory networks}.
\bjtitle{Neurocomputing}
\bvolume{469},
\bfpage{198}--\blpage{208}
(\byear{2022})
\end{barticle}
\endbibitem

%%% 45
\bibitem[\protect\citeauthoryear{McEliece et~al.}{1987}]{mceliece1987capacity}
\begin{barticle}
\bauthor{\bsnm{McEliece}, \binits{R.}},
\bauthor{\bsnm{Posner}, \binits{E.}},
\bauthor{\bsnm{Rodemich}, \binits{E.}},
\bauthor{\bsnm{Venkatesh}, \binits{S.}}:
\batitle{The capacity of the hopfield associative memory}.
\bjtitle{IEEE transactions on Information Theory}
\bvolume{33}(\bissue{4}),
\bfpage{461}--\blpage{482}
(\byear{1987})
\end{barticle}
\endbibitem

%%% 46
\bibitem[\protect\citeauthoryear{De~Marzo and Iannelli}{2023}]{de2023effect}
\begin{barticle}
\bauthor{\bsnm{De~Marzo}, \binits{G.}},
\bauthor{\bsnm{Iannelli}, \binits{G.}}:
\batitle{Effect of spatial correlations on hopfield neural network and dense associative memories}.
\bjtitle{Physica A: Statistical Mechanics and its Applications}
\bvolume{612},
\bfpage{128487}
(\byear{2023})
\end{barticle}
\endbibitem

%%% 47
\bibitem[\protect\citeauthoryear{Zhou et~al.}{2022}]{zhou2022photonic}
\begin{barticle}
\bauthor{\bsnm{Zhou}, \binits{H.}},
\bauthor{\bsnm{Dong}, \binits{J.}},
\bauthor{\bsnm{Cheng}, \binits{J.}},
\bauthor{\bsnm{Dong}, \binits{W.}},
\bauthor{\bsnm{Huang}, \binits{C.}},
\bauthor{\bsnm{Shen}, \binits{Y.}},
\bauthor{\bsnm{Zhang}, \binits{Q.}},
\bauthor{\bsnm{Gu}, \binits{M.}},
\bauthor{\bsnm{Qian}, \binits{C.}},
\bauthor{\bsnm{Chen}, \binits{H.}}, \betal:
\batitle{Photonic matrix multiplication lights up photonic accelerator and beyond}.
\bjtitle{Light: Science \& Applications}
\bvolume{11}(\bissue{1}),
\bfpage{30}
(\byear{2022})
\end{barticle}
\endbibitem

%%% 48
\bibitem[\protect\citeauthoryear{Fang et~al.}{2021}]{fang2021experimental}
\begin{barticle}
\bauthor{\bsnm{Fang}, \binits{Y.}},
\bauthor{\bsnm{Huang}, \binits{J.}},
\bauthor{\bsnm{Ruan}, \binits{Z.}}:
\batitle{Experimental observation of phase transitions in spatial photonic ising machine}.
\bjtitle{Physical Review Letters}
\bvolume{127}(\bissue{4}),
\bfpage{043902}
(\byear{2021})
\end{barticle}
\endbibitem

%%% 49
\bibitem[\protect\citeauthoryear{Kumar et~al.}{2020}]{kumar2020large}
\begin{barticle}
\bauthor{\bsnm{Kumar}, \binits{S.}},
\bauthor{\bsnm{Zhang}, \binits{H.}},
\bauthor{\bsnm{Huang}, \binits{Y.-P.}}:
\batitle{Large-scale ising emulation with four body interaction and all-to-all connections}.
\bjtitle{Communications Physics}
\bvolume{3}(\bissue{1}),
\bfpage{108}
(\byear{2020})
\end{barticle}
\endbibitem

%%% 50
\bibitem[\protect\citeauthoryear{Pierangeli et~al.}{2019}]{pierangeli2019large}
\begin{barticle}
\bauthor{\bsnm{Pierangeli}, \binits{D.}},
\bauthor{\bsnm{Marcucci}, \binits{G.}},
\bauthor{\bsnm{Conti}, \binits{C.}}:
\batitle{Large-scale photonic ising machine by spatial light modulation}.
\bjtitle{Physical review letters}
\bvolume{122}(\bissue{21}),
\bfpage{213902}
(\byear{2019})
\end{barticle}
\endbibitem

%%% 51
\bibitem[\protect\citeauthoryear{Wang et~al.}{2024}]{wang2024large}
\begin{barticle}
\bauthor{\bsnm{Wang}, \binits{H.}},
\bauthor{\bsnm{Hu}, \binits{J.}},
\bauthor{\bsnm{Morandi}, \binits{A.}},
\bauthor{\bsnm{Nardi}, \binits{A.}},
\bauthor{\bsnm{Xia}, \binits{F.}},
\bauthor{\bsnm{Li}, \binits{X.}},
\bauthor{\bsnm{Savo}, \binits{R.}},
\bauthor{\bsnm{Liu}, \binits{Q.}},
\bauthor{\bsnm{Grange}, \binits{R.}},
\bauthor{\bsnm{Gigan}, \binits{S.}}:
\batitle{Large-scale photonic computing with nonlinear disordered media}.
\bjtitle{Nature Computational Science}
\bvolume{4}(\bissue{6}),
\bfpage{429}--\blpage{439}
(\byear{2024})
\end{barticle}
\endbibitem

%%% 52
\bibitem[\protect\citeauthoryear{Hedayat and Wallis}{1978}]{hedayat1978hadamard}
\begin{botherref}
\oauthor{\bsnm{Hedayat}, \binits{A.}},
\oauthor{\bsnm{Wallis}, \binits{W.D.}}:
Hadamard matrices and their applications.
The annals of statistics,
1184--1238
(1978)
\end{botherref}
\endbibitem

%%% 53
\bibitem[\protect\citeauthoryear{Kumar et~al.}{2023}]{kumar2023observation}
\begin{barticle}
\bauthor{\bsnm{Kumar}, \binits{S.}},
\bauthor{\bsnm{Li}, \binits{Z.}},
\bauthor{\bsnm{Bu}, \binits{T.}},
\bauthor{\bsnm{Qu}, \binits{C.}},
\bauthor{\bsnm{Huang}, \binits{Y.}}:
\batitle{Observation of distinct phase transitions in a nonlinear optical ising machine}.
\bjtitle{Communications Physics}
\bvolume{6}(\bissue{1}),
\bfpage{31}
(\byear{2023})
\end{barticle}
\endbibitem

%%% 54
\bibitem[\protect\citeauthoryear{Yamashita et~al.}{2023}]{yamashita2023low}
\begin{barticle}
\bauthor{\bsnm{Yamashita}, \binits{H.}},
\bauthor{\bsnm{Okubo}, \binits{K.-i.}},
\bauthor{\bsnm{Shimomura}, \binits{S.}},
\bauthor{\bsnm{Ogura}, \binits{Y.}},
\bauthor{\bsnm{Tanida}, \binits{J.}},
\bauthor{\bsnm{Suzuki}, \binits{H.}}:
\batitle{Low-rank combinatorial optimization and statistical learning by spatial photonic ising machine}.
\bjtitle{Physical Review Letters}
\bvolume{131}(\bissue{6}),
\bfpage{063801}
(\byear{2023})
\end{barticle}
\endbibitem

%%% 55
\bibitem[\protect\citeauthoryear{Yamamoto et~al.}{2017}]{yamamoto2017time}
\begin{bchapter}
\bauthor{\bsnm{Yamamoto}, \binits{K.}},
\bauthor{\bsnm{Huang}, \binits{W.}},
\bauthor{\bsnm{Takamaeda-Yamazaki}, \binits{S.}},
\bauthor{\bsnm{Ikebe}, \binits{M.}},
\bauthor{\bsnm{Asai}, \binits{T.}},
\bauthor{\bsnm{Motomura}, \binits{M.}}:
\bctitle{A time-division multiplexing ising machine on fpgas}.
In: \bbtitle{Proceedings of the 8th International Symposium on Highly Efficient Accelerators and Reconfigurable Technologies},
pp. \bfpage{1}--\blpage{6}
(\byear{2017})
\end{bchapter}
\endbibitem

%%% 56
\bibitem[\protect\citeauthoryear{Deng}{2012}]{6296535}
\begin{barticle}
\bauthor{\bsnm{Deng}, \binits{L.}}:
\batitle{The mnist database of handwritten digit images for machine learning research [best of the web]}.
\bjtitle{IEEE Signal Processing Magazine}
\bvolume{29}(\bissue{6}),
\bfpage{141}--\blpage{142}
(\byear{2012})
\doiurl{10.1109/MSP.2012.2211477}
\end{barticle}
\endbibitem

%%% 57
\bibitem[\protect\citeauthoryear{Ye et~al.}{2021}]{ye2021high}
\begin{barticle}
\bauthor{\bsnm{Ye}, \binits{X.}},
\bauthor{\bsnm{Ni}, \binits{F.}},
\bauthor{\bsnm{Li}, \binits{H.}},
\bauthor{\bsnm{Liu}, \binits{H.}},
\bauthor{\bsnm{Zheng}, \binits{Y.}},
\bauthor{\bsnm{Chen}, \binits{X.}}:
\batitle{High-speed programmable lithium niobate thin film spatial light modulator}.
\bjtitle{Optics Letters}
\bvolume{46}(\bissue{5}),
\bfpage{1037}--\blpage{1040}
(\byear{2021})
\end{barticle}
\endbibitem

%%% 58
\bibitem[\protect\citeauthoryear{Mitchell et~al.}{2016}]{mitchell2016high}
\begin{barticle}
\bauthor{\bsnm{Mitchell}, \binits{K.J.}},
\bauthor{\bsnm{Turtaev}, \binits{S.}},
\bauthor{\bsnm{Padgett}, \binits{M.J.}},
\bauthor{\bsnm{{\v{C}}i{\v{z}}m{\'a}r}, \binits{T.}},
\bauthor{\bsnm{Phillips}, \binits{D.B.}}:
\batitle{High-speed spatial control of the intensity, phase and polarisation of vector beams using a digital micro-mirror device}.
\bjtitle{Optics express}
\bvolume{24}(\bissue{25}),
\bfpage{29269}--\blpage{29282}
(\byear{2016})
\end{barticle}
\endbibitem

%%% 59
\bibitem[\protect\citeauthoryear{Smolyaninov et~al.}{2019}]{smolyaninov2019programmable}
\begin{barticle}
\bauthor{\bsnm{Smolyaninov}, \binits{A.}},
\bauthor{\bsnm{El~Amili}, \binits{A.}},
\bauthor{\bsnm{Vallini}, \binits{F.}},
\bauthor{\bsnm{Pappert}, \binits{S.}},
\bauthor{\bsnm{Fainman}, \binits{Y.}}:
\batitle{Programmable plasmonic phase modulation of free-space wavefronts at gigahertz rates}.
\bjtitle{Nature Photonics}
\bvolume{13}(\bissue{6}),
\bfpage{431}--\blpage{435}
(\byear{2019})
\end{barticle}
\endbibitem

%%% 60
\bibitem[\protect\citeauthoryear{Liu et~al.}{2017}]{liu2017focusing}
\begin{barticle}
\bauthor{\bsnm{Liu}, \binits{Y.}},
\bauthor{\bsnm{Ma}, \binits{C.}},
\bauthor{\bsnm{Shen}, \binits{Y.}},
\bauthor{\bsnm{Shi}, \binits{J.}},
\bauthor{\bsnm{Wang}, \binits{L.V.}}:
\batitle{Focusing light inside dynamic scattering media with millisecond digital optical phase conjugation}.
\bjtitle{Optica}
\bvolume{4}(\bissue{2}),
\bfpage{280}--\blpage{288}
(\byear{2017})
\end{barticle}
\endbibitem

%%% 61
\bibitem[\protect\citeauthoryear{Veraldi et~al.}{2025}]{veraldi2025fully}
\begin{barticle}
\bauthor{\bsnm{Veraldi}, \binits{D.}},
\bauthor{\bsnm{Pierangeli}, \binits{D.}},
\bauthor{\bsnm{Gentilini}, \binits{S.}},
\bauthor{\bsnm{Strinati}, \binits{M.C.}},
\bauthor{\bsnm{Sakellariou}, \binits{J.}},
\bauthor{\bsnm{Cummins}, \binits{J.S.}},
\bauthor{\bsnm{Kamaletdinov}, \binits{A.}},
\bauthor{\bsnm{Syed}, \binits{M.}},
\bauthor{\bsnm{Wang}, \binits{R.Z.}},
\bauthor{\bsnm{Berloff}, \binits{N.G.}}, \betal:
\batitle{Fully programmable spatial photonic ising machine by focal plane division}.
\bjtitle{Physical Review Letters}
\bvolume{134}(\bissue{6}),
\bfpage{063802}
(\byear{2025})
\end{barticle}
\endbibitem

%%% 62
\bibitem[\protect\citeauthoryear{Luo et~al.}{2023}]{luo2023wavelength}
\begin{barticle}
\bauthor{\bsnm{Luo}, \binits{L.}},
\bauthor{\bsnm{Mi}, \binits{Z.}},
\bauthor{\bsnm{Huang}, \binits{J.}},
\bauthor{\bsnm{Ruan}, \binits{Z.}}:
\batitle{Wavelength-division multiplexing optical ising simulator enabling fully programmable spin couplings and external magnetic fields}.
\bjtitle{Science Advances}
\bvolume{9}(\bissue{48}),
\bfpage{6238}
(\byear{2023})
\end{barticle}
\endbibitem

%%% 63
\bibitem[\protect\citeauthoryear{Sakabe et~al.}{2023}]{sakabe2023spatial}
\begin{barticle}
\bauthor{\bsnm{Sakabe}, \binits{T.}},
\bauthor{\bsnm{Shimomura}, \binits{S.}},
\bauthor{\bsnm{Ogura}, \binits{Y.}},
\bauthor{\bsnm{Okubo}, \binits{K.-i.}},
\bauthor{\bsnm{Yamashita}, \binits{H.}},
\bauthor{\bsnm{Suzuki}, \binits{H.}},
\bauthor{\bsnm{Tanida}, \binits{J.}}:
\batitle{Spatial-photonic ising machine by space-division multiplexing with physically tunable coefficients of a multi-component model}.
\bjtitle{Optics Express}
\bvolume{31}(\bissue{26}),
\bfpage{44127}--\blpage{44138}
(\byear{2023})
\end{barticle}
\endbibitem

%%% 64
\bibitem[\protect\citeauthoryear{Fiorelli et~al.}{2019}]{fiorelli2019quantum}
\begin{barticle}
\bauthor{\bsnm{Fiorelli}, \binits{E.}},
\bauthor{\bsnm{Rotondo}, \binits{P.}},
\bauthor{\bsnm{Marcuzzi}, \binits{M.}},
\bauthor{\bsnm{Garrahan}, \binits{J.P.}},
\bauthor{\bsnm{Lesanovsky}, \binits{I.}}:
\batitle{Quantum accelerated approach to the thermal state of classical all-to-all connected spin systems with applications to pattern retrieval in the hopfield neural network}.
\bjtitle{Physical Review A}
\bvolume{99}(\bissue{3}),
\bfpage{032126}
(\byear{2019})
\end{barticle}
\endbibitem

%%% 65
\bibitem[\protect\citeauthoryear{Brunner et~al.}{2025}]{brunner2025roadmap}
\begin{botherref}
\oauthor{\bsnm{Brunner}, \binits{D.}},
\oauthor{\bsnm{Shastri}, \binits{B.J.}},
\oauthor{\bsnm{Qadasi}, \binits{M.A.A.}},
\oauthor{\bsnm{Ballani}, \binits{H.}},
\oauthor{\bsnm{Barbay}, \binits{S.}},
\oauthor{\bsnm{Biasi}, \binits{S.}},
\oauthor{\bsnm{Bienstman}, \binits{P.}},
\oauthor{\bsnm{Bilodeau}, \binits{S.}},
\oauthor{\bsnm{Bogaerts}, \binits{W.}},
\oauthor{\bsnm{B{\"o}hm}, \binits{F.}}, et al.:
Roadmap on neuromorphic photonics.
arXiv preprint arXiv:2501.07917
(2025)
\end{botherref}
\endbibitem

%%% 66
\bibitem[\protect\citeauthoryear{Jordan and Mitchell}{2015}]{jordan2015machine}
\begin{barticle}
\bauthor{\bsnm{Jordan}, \binits{M.I.}},
\bauthor{\bsnm{Mitchell}, \binits{T.M.}}:
\batitle{Machine learning: Trends, perspectives, and prospects}.
\bjtitle{Science}
\bvolume{349}(\bissue{6245}),
\bfpage{255}--\blpage{260}
(\byear{2015})
\end{barticle}
\endbibitem

%%% 67
\bibitem[\protect\citeauthoryear{Musa}{2025}]{Musa_Correlated_Dense_Associative_2025}
\begin{botherref}
\oauthor{\bsnm{Musa}, \binits{K.}}:
{Correlated Dense Associative Memory OHNN}
(2025).
\url{https://github.com/kmusa11/OHNN-CDAM.git}
\end{botherref}
\endbibitem

%%% 68
\bibitem[\protect\citeauthoryear{Wu and Xu}{2010}]{wu2010correlation}
\begin{bchapter}
\bauthor{\bsnm{Wu}, \binits{W.-J.}},
\bauthor{\bsnm{Xu}, \binits{Y.}}:
\bctitle{Correlation analysis of visual verbs' subcategorization based on pearson's correlation coefficient}.
In: \bbtitle{2010 International Conference on Machine Learning and Cybernetics},
vol. \bseriesno{4},
pp. \bfpage{2042}--\blpage{2046}
(\byear{2010}).
\bcomment{IEEE}
\end{bchapter}
\endbibitem

%%% 69
\bibitem[\protect\citeauthoryear{Paley}{1933}]{sapm1933121311}
\begin{barticle}
\bauthor{\bsnm{Paley}, \binits{R.E.A.C.}}:
\batitle{On orthogonal matrices}.
\bjtitle{Journal of Mathematics and Physics}
\bvolume{12}(\bissue{1-4}),
\bfpage{311}--\blpage{320}
(\byear{1933})
\doiurl{10.1002/sapm1933121311}
{\href{https://arxiv.org/abs/https://onlinelibrary.wiley.com/doi/pdf/10.1002/sapm1933121311}{{https://onlinelibrary.wiley.com/doi/pdf/10.1002/sapm1933121311}}}
\end{barticle}
\endbibitem

%%% 70
\bibitem[\protect\citeauthoryear{Jones}{2017}]{jones2017paleypaleygraphs}
\begin{botherref}
\oauthor{\bsnm{Jones}, \binits{G.A.}}:
Paley and the Paley graphs
(2017).
\url{https://arxiv.org/abs/1702.00285}
\end{botherref}
\endbibitem

%%% 71
\bibitem[\protect\citeauthoryear{Caprara et~al.}{2014}]{Caprara2014GenerationOA}
\begin{barticle}
\bauthor{\bsnm{Caprara}, \binits{A.}},
\bauthor{\bsnm{Furini}, \binits{F.}},
\bauthor{\bsnm{Lodi}, \binits{A.}},
\bauthor{\bsnm{Mangia}, \binits{M.}},
\bauthor{\bsnm{Rovatti}, \binits{R.}},
\bauthor{\bsnm{Setti}, \binits{G.}}:
\batitle{Generation of antipodal random vectors with prescribed non-stationary 2-nd order statistics}.
\bjtitle{IEEE Transactions on Signal Processing}
\bvolume{62},
\bfpage{1603}--\blpage{1612}
(\byear{2014})
\end{barticle}
\endbibitem

%%% 72
\bibitem[\protect\citeauthoryear{Bradley}{2015}]{mathoverflow_bernoulli}
\begin{botherref}
\oauthor{\bsnm{Bradley}, \binits{B.}}:
Generate Bernoulli vector with given covariance matrix.
MathOverflow
(2015).
\url{https://mathoverflow.net/questions/210483/generate-bernoulli-vector-with-given-covariance-matrix}
\end{botherref}
\endbibitem

\end{thebibliography}

\end{document}